\documentclass[aps,prl,twocolumn,superscriptaddress,floatfix,nofootinbib]{revtex4-1}
\usepackage[english]{babel}
\usepackage[T1]{fontenc}
\usepackage[colorlinks=true,linkcolor=blue,citecolor=blue,urlcolor=blue]{hyperref}
\usepackage{bm,bbm,amssymb,amsmath}
\usepackage{graphicx}
\usepackage{multirow,booktabs,dcolumn}
\usepackage{slashed}
\usepackage{isotope}
\usepackage{physics}
\usepackage[capitalize]{cleveref}
\usepackage{nicefrac}
\usepackage{ulem}

\begin{document}

\title{Nuclear and neutron-star matter from local chiral interactions}

\author{D. Lonardoni}
\affiliation{Facility for Rare Isotope Beams, Michigan State University, East Lansing, Michigan 48824, USA}
\affiliation{Theoretical Division, Los Alamos National Laboratory, Los Alamos, New Mexico 87545, USA}

\author{I. Tews}
\affiliation{Theoretical Division, Los Alamos National Laboratory, Los Alamos, New Mexico 87545, USA}

\author{S. Gandolfi}
\affiliation{Theoretical Division, Los Alamos National Laboratory, Los Alamos, New Mexico 87545, USA}

\author{J. Carlson}
\affiliation{Theoretical Division, Los Alamos National Laboratory, Los Alamos, New Mexico 87545, USA}

\begin{abstract}
We report a quantum Monte Carlo calculation of the equation
of state of symmetric nuclear matter using local interactions derived
from chiral effective field theory up to next-to-next-to-leading order 
fit to few-body observables only. The empirical saturation
density and energy are well reproduced within statistical and
systematic uncertainties. We have also derived the symmetry energy as a
function of the density, finding good agreement with available experimentally derived constraints
at saturation and twice saturation density. We find that the corresponding pressure is also
in excellent agreement with recent constraints extracted from gravitational waves of 
the neutron-star merger GW170817 by the LIGO-Virgo detection.
\end{abstract}

\maketitle

\textit{Introduction.} 
The nuclear equation of state (EOS)
is of great interest for nuclear physics and nuclear
astrophysics. At proton fractions $x\sim 0.5$, for the so-called
symmetric nuclear matter (SNM), the EOS sets the bulk properties of
atomic nuclei and determines where atomic nuclei saturate. At lower
proton fractions, $x\lesssim 0.1$, the nuclear EOS determines the
properties of neutron stars.
The energy difference of nuclear matter at different proton fractions,
e.g., between SNM and pure neutron matter (PNM) with $x=0$, is governed
by the nuclear symmetry energy $S(n)$, with $n$ the baryon density.
The nuclear symmetry energy is a fundamental physical quantity that
affects a range of neutron-star properties, such as cooling rates,
the thickness of the neutron-star crust, the mass-radius 
relation~\cite{Hebeler:2010jx,Gandolfi:2012,Steiner:2012},
and the moment of inertia~\cite{Tsang:2012,lattimer14}, and is
deeply connected to properties of atomic nuclei, e.g., the dipole
polarizability, the giant dipole resonance, and the neutron skin of
neutron-rich nuclei~\cite{Lattimer:2012xj}. Hence, it is possible
to infer properties of the nuclear EOS at low densities and larger
proton fractions from laboratory experiments with atomic nuclei, e.g.,
in the future Facility for Rare Isotope Beams (FRIB) at Michigan State
University. The very neutron-rich nuclear matter that forms neutron
stars, on the other hand, cannot be directly probed in any terrestrial
experiment. Hence, neutron stars provide a unique and complementary
laboratory for dense nuclear matter at extreme conditions.

Understanding the properties of the nuclear EOS has recently
become even more critical with the advent of gravitational-wave
(GW) astronomy and the first-ever detection of gravitational waves
from a binary neutron-star merger by the LIGO-Virgo collaboration in
2017~\cite{TheLIGOScientific:2017qsa,GBM:2017lvd}. This GW detection,
together with the identification of such mergers as a site of $r$-process
nucleosynthesis, has attracted tremendous interest. Many expected future
observations during the current and future LIGO observing runs, like the
first detection of a neutron-star black-hole merger event (S190814bv),
will be of similar high impact. Reliable and precise calculations
of the nuclear EOS in neutron-rich regimes and for sufficiently high
densities are needed to accurately extract information from such
astrophysical observations. In addition, reliable calculations of the
nuclear EOS at various proton fractions are needed in order to compare to
experiments extracting the symmetry energy. Such calculations require a
systematic theory for strong interactions that enables estimations of theoretical
uncertainties, in combination with advanced many-body methods to solve
the complicated nuclear many-body problem.

In this work, we address this timely, high-impact topic, and present 
quantum Monte Carlo (QMC) calculations of both neutron and
symmetric nuclear matter using local chiral effective field theory (EFT)
interactions~\cite{Gezerlis:2014,Lynn:2016}, in order to constrain the
nuclear EOS and the symmetry energy with robust uncertainty estimates. 
Our goal is to compare QMC results directly to observational/experimental 
constraints, rather than to other many-body calculations of nuclear 
matter~\cite{Akmal:1998,Hebeler:2010xb,Hagen:2014,Ekstrom:2015,Holt:2016pjb,Drischler:2017wtt,Carbone:2019}, 
which will require a more detailed discussion on methods and interactions, 
outside the scope of this work.

\textit{Hamiltonian and method.} 
We describe the nuclear many-body system as a collection of
pointlike interacting nucleons with a nonrelativistic
Hamiltonian $H$ that includes two-body $(v_{ij})$ and
three-body $(V_{ijk})$ potentials~\cite{Carlson:2015}. In
this work we employ local interactions derived from chiral
EFT up to next-to-next-to-leading order
(N$^2$LO)~\cite{Gezerlis:2013,Gezerlis:2014,Lynn:2016}.
These give a reasonable description of nucleon-nucleon
phase shifts within uncertainty estimates up to laboratory energies of
$500\,\rm MeV$, except in the triplet $D$ waves.
They include consistent two- and three-body
potentials, and have been used to study the ground-state properties of
nuclei~\cite{Lynn:2016,Lynn:2017,Lonardoni:2018prl,Lonardoni:2018prc},
few-neutron
systems~\cite{Gandolfi:2016,Gandolfi:2017}, and neutron-star
matter~\cite{Gezerlis:2014,Tews:2016,Lynn:2016,Buraczynski:2016,Tews:2018apj,Tews:2018,Tews:2019}.
In particular, we consider the interactions with coordinate-space cutoff $R_0=1.0\,\rm fm$ and 
two different parametrizations of the three-body contact term $V_E$, namely $E\mathbbm1$ and $E\tau$. 
$E\mathbbm1$ employs the identity operator $\mathbbm{1}$ between particle $i$ and $j$, 
while $E\tau$ employs the isospin operator structure $\bm\tau_i\cdot\bm\tau_j$.
The difference of the two parametrizations is due to regulator and cutoff artifacts. 
A comparison of both interactions allows one to quantify the uncertainty due to the regularization scheme and scale.
We do not consider the cutoff $R_0=1.2\,\rm fm$ because it leads to 
severe regulator artifacts in atomic nuclei~\cite{Lonardoni:2018prc}.
For local chiral interactions, 
the short-range contact term $V_E$, together with the three-body one-pion-exchange--contact term 
$V_D$, are characterized by two low-energy couplings $c_D$ and $c_E$ which are fit to few-body observables, 
namely the $\alpha$-particle binding energy and
the spin-orbit splitting in the neutron-$\alpha$ $P$-wave phase
shifts (see Refs.~\cite{Lynn:2016,Lonardoni:2018prc} for details). 
However, the interactions are capable of describing ground-state
properties of nuclei up to (at least) $A=16$~\cite{Lonardoni:2018prl},
and the $E\mathbbm1$ potential can simultaneously predict
properties of neutron matter compatible with astrophysical
observations~\cite{Tews:2018apj}.

The starting point of all QMC methods is the choice of a wave
function representing the system, typically expressed as the trial
state $|\Psi_T\rangle=\mathcal F\,|\Phi\rangle$, where $\mathcal
F$ is the correlation operator acting between pairs and triplets
of particles. $\mathcal F$ incorporates strong spin and isospin
dependence into the trial state, as induced by the employed nuclear
Hamiltonian. For infinite matter, the term $|\Phi\rangle$ is built
from a Slater determinant of plane waves $\phi_{\bm k}(i)=e^{i\bm
k\cdot\bm r_i}$ with momenta discretized in a finite box whose volume
is determined by the chosen baryon density $n$ and number of particles
$A$~\cite{Gandolfi:2014}. The infinite system is then realized by
applying periodic boundary conditions~\cite{Gandolfi:2009}. Finite-size
corrections to the energy results are included as described in
Refs.~\cite{Sarsa:2003,Gandolfi:2014}. All the parameters of $\Psi_T$
are chosen by minimizing the variational energy as described in
Ref.~\cite{Sorella:2001}.

In this work, we make use of the auxiliary field diffusion Monte Carlo
(AFDMC) method~\cite{Schmidt:1999,Carlson:2015,Lonardoni:2018prc} that
allows one to access the true ground state of a nuclear system by
evolving the initial trial state in imaginary time according to the
projection operator $\exp[-(H-E_T)\tau]|\Psi_T\rangle$, where $E_T$ is
a normalization parameter.  In the limit of infinite imaginary time,
higher-energy components in the trial state are filtered out, and the system is
projected onto the ground state.  
Such imaginary-time evolution is performed by sampling both spatial coordinates and
spin/isosospin configurations, the latter via a Hubbard-Stratonovich
transformation. For local chiral interactions, the propagation of
three-body terms is carried out via an effective Hamiltonian as described
in Refs.~\cite{Lonardoni:2018prl,Lonardoni:2018prc}.
The sign problem is initially suppressed by evolving the trial wave function using the
constrained-path approximation~\cite{Zhang:2003}.  An unconstrained
evolution is then performed until the sign problem dominates and the
variance of the results becomes severely large.  Finally, expectation
values are evaluated over the sampled configurations to compute the
relevant observables~\cite{Lonardoni:2018prc}.
Note that, similarly to the case of atomic nuclei, the approximate 
propagation of three-body terms and the unconstrained evolution for
infinite matter are robust and under control. The associated uncertainties 
are included in the final Monte Carlo uncertainty estimate.

The AFDMC method has been used in the past to
calculate the EOS of PNM using both phenomenological and local chiral
interactions~\cite{Gandolfi:2009,Gandolfi:2012,Gandolfi:2015,Tews:2016,Lynn:2016,Buraczynski:2016,Tews:2018apj,Tews:2018,Tews:2019,Piarulli:2019},
and attempts at calculating the EOS of asymmetric nuclear matter have been
carried out for a simplified phenomenological model~\cite{Gandolfi:2014}.
Here, we perform a study of the EOS of PNM and
SNM using local chiral interactions up to N$^2$LO. We consider,
respectively, 66 neutrons and 28 nucleons in a periodic box described
by the trial state $|\Psi_T\rangle$, where the spin/isospin-dependent
two- and three-body
correlations are expressed as a sum of linear and quadratic spin/isospin
operators as in Ref.~\cite{Lonardoni:2018prc}.
Note that finite size effects in SNM are smaller than in PNM~\cite{Gandolfi:2014},
hence, a smaller particle number is sufficient.

Due to the high computational cost of performing derivatives of
the trial wave function, no spin-orbit correlations are 
typically included in the AFDMC wave function.
However, as reported in Ref.~\cite{Brualla:2003}, the
largest contribution to the total energy given by spin-orbit terms can
be obtained by using a simplified spin-orbit correlation that can be
implemented in the AFDMC wave function 
by substituting the plane wave $e^{i\bm k\cdot\bm r_i}$ with
\begin{align}
    \phi_{\bm k}(i)=\exp \Bigg({i\bm k\cdot\bm r_i+\frac{\beta}{2}\sum_{j\ne i}f_{ij}^{\rm LS}\,\bm r_{ij}\cross\bm k\cdot\bm\sigma_i}\Bigg),
	\label{eq:sbf}
\end{align}
where $\beta$ is a variational parameter, and $f_{ij}^{\rm LS}$ is 
a spin-orbit radial function obtained from the solution of Schr\"odinger-like
equations in the relative distance $r_{ij}$ as described in Ref.~\cite{Carlson:2015}.
\Cref{eq:sbf} defines the so-called spin-backflow correlations, 
and is analogous to the implementation of standard backflow correlations 
in fermionic systems~\cite{Schmidt:1981}. Spin-backflow correlations only
imply spin rotations among the components of the Slater determinant
$|\Phi\rangle$~\cite{Brualla:2003}, which makes them computationally
cheap.
As shown in a simplified case for PNM in Ref.~\cite{Brualla:2003},
such correlations greatly improve the quality of the wave function,
and their contribution to the total energy is not negligible.

\begin{figure*}[t]
\includegraphics[width=0.5\linewidth]{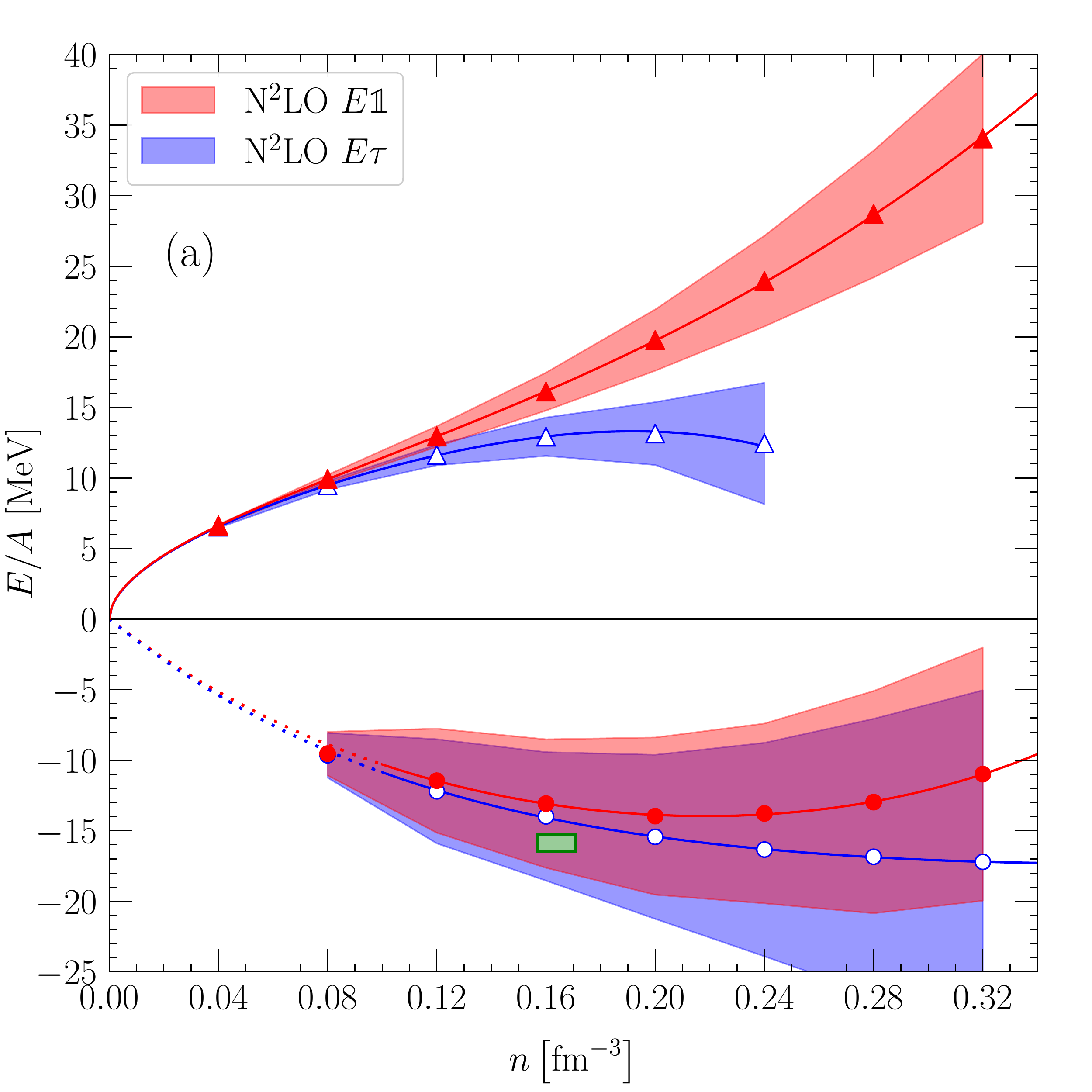}\hfill
\includegraphics[width=0.5\linewidth]{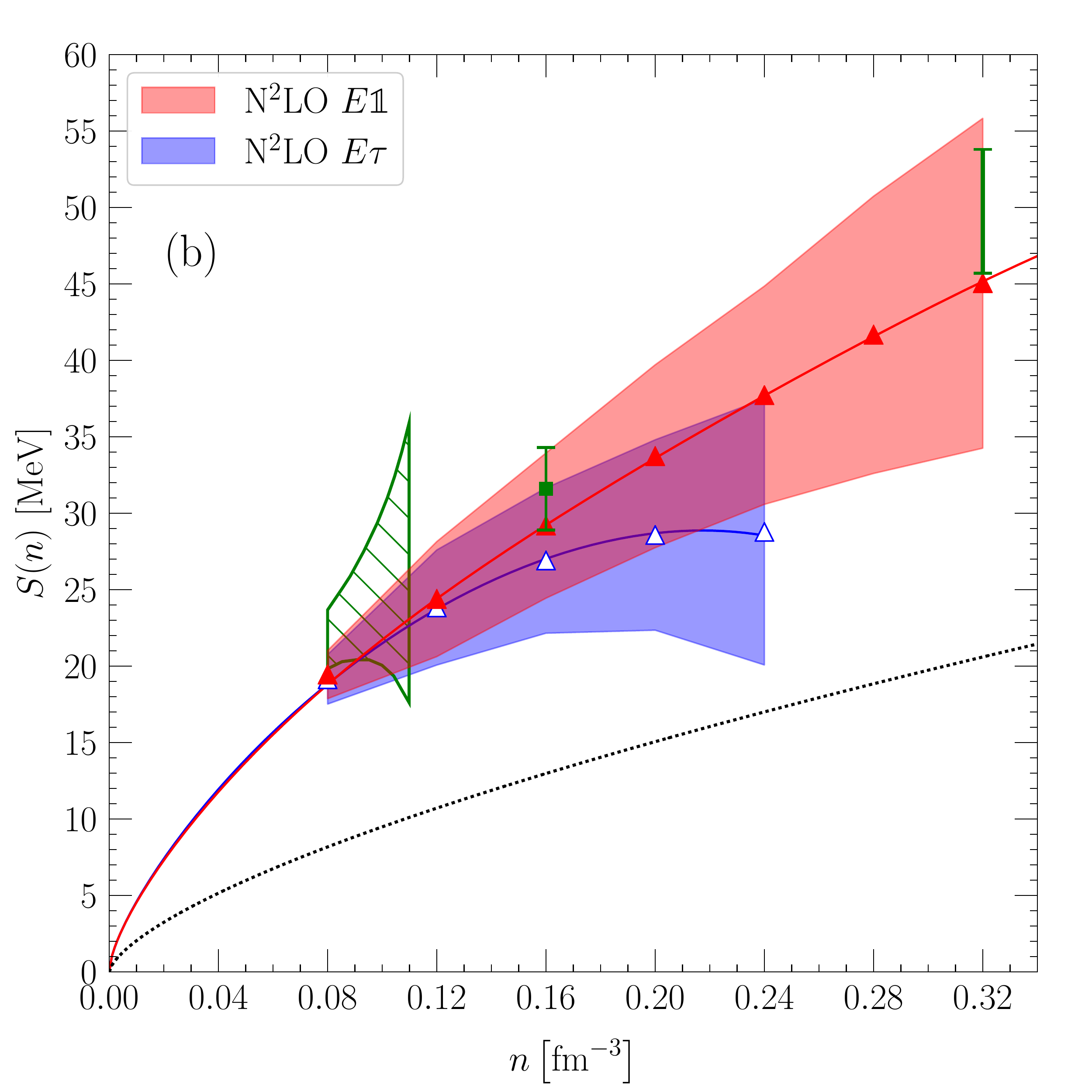}
\caption[]{(a) Equation of sate of PNM (triangles) and SNM (circles). (b) Nuclear symmetry energy. Red solid symbols (blue empty symbols) are AFDMC results for the N$^2$LO local chiral potentials with coordinate-space cutoff $R_0=1.0\,\rm fm$ and the $E\mathbbm1$ ($E\tau$) parametrization of the three-body force. The EOS curves are fit to AFDMC results using \cref{eq:pnm,eq:snm} (see parameters in \cref{tab:par}). Colored bands represent the uncertainties of the many-body calculations, which include both statistical Monte Carlo errors and the uncertainties coming from the truncation of the chiral expansion. In panel (a), the green box indicates the empirical saturation point~\cite{Drischler:2017wtt}. In panel (b), we show experimental constraints on the symmetry energy below saturation density from Ref.~\cite{Zhang:2015}, at saturation~\cite{Li:2019} and twice saturation density [for $S(n_{\rm sat})=31\,\rm MeV$]~\cite{Russotto:2016}. The dashed black curve is the Fermi gas result.}
\label{fig:eos-esym}
\end{figure*}

\textit{Results.} 
The AFDMC results for the EOS of PNM and SNM at
N$^2$LO are shown in the left panel of \cref{fig:eos-esym} for the
$E\mathbbm1$ (red) and $E\tau$ (blue) parametrizations. A
table with the energy results as a function of the density is available
in the Supplemental Material~\cite{prr:supp}.

We have found that spin/isospin-dependent correlations
yield a negligible improvement of the total energy in PNM, while, similarly
to the case of atomic nuclei~\cite{Lonardoni:2018prc}, they have a large
effect in SNM, where tensor contributions are much stronger.
For a given momentum scale, spin-backflow correlations are very effective and have similar contributions in both PNM and SNM, with an energy gain of
the order of $-0.5\,{\rm MeV}/A$ at the momentum scale $k\simeq1.33\,\rm fm^{-1}$.
As done in Ref.~\cite{Piarulli:2019}, AFDMC results for PNM are obtained at the
constrained-path level for 66 neutrons and successively corrected with
the energy gain obtained from the unconstrained evolution of 14 neutrons.
AFDMC results for SNM are obtained 
at the constrained-path level only. 
In this case in fact, the unconstrained evolution
indicates minimal energy corrections,
of the order of $<2\%$ at saturation density,
but requires large computational resources.

The solid curves in \cref{fig:eos-esym} are fit to the AFDMC results
according to the relations~\cite{Margueron:2017,Gandolfi:2019zpj}
\begin{align}
    E_{\rm PNM}(n)=a\left(\frac{n}{n_{\rm sat}}\right)^\alpha+b\left(\frac{n}{n_{\rm sat}}\right)^\beta, 
    \label{eq:pnm}
\end{align}
\vspace{-0.45cm}
\begin{align}
    E_{\rm SNM}(n)=&E_0+\frac{K_0}{2!}\left(\frac{n-n_0}{3n_0}\right)^2+\frac{Q_0}{3!}\left(\frac{n-n_0}{3n_0}\right)^3 \nonumber \\ &+\frac{Z_0}{4!}\left(\frac{n-n_0}{3n_0}\right)^4 +O\left(\frac{n-n_0}{3n_0}\right)^5 ,
    \label{eq:snm}
\end{align}
where $n_{\rm sat}=0.16\,\rm fm^{-3}$ is the empirical saturation density,
$n_0$ and $E_0$ are saturation density and saturation energy for the
given Hamiltonian, and $K_0$, $Q_0$, and $Z_0$ are the incompressibility,
skewness, and kurtosis parameters.  For SNM we fit the AFDMC energies
above $n=0.12\,\rm fm^{-3}$, since clustering is expected to appear
at lower densities.  Assuming that the system behaves as
uniform matter over the whole density range, which is not a realistic
picture for low density nuclear matter (hence the dashed curves in
\cref{fig:eos-esym}), we enforce $E_{\rm SNM}(0)=0$ by adjusting $Z_0$
accordingly. All the fitting parameters, together with the empirical
values, where available, are reported in \cref{tab:par}.

\begin{table}[b]
\centering
\caption[]{Fitting parameters for \cref{eq:pnm,eq:snm}, where the
errors originate in the statistical Monte Carlo uncertainties only. 
Empirical values from Refs.~\cite{Drischler:2017wtt,Margueron:2017} are shown for comparison.}
\label{tab:par}
\begin{ruledtabular}
\begin{tabular}{cccc}
Par. & N$^2$LO$\,E\mathbbm1$ & N$^2$LO$\,E\tau$ & Empirical\\
\midrule
$a$      & $13.9(2)\,\rm MeV$ & $13.9(3)\,\rm MeV$ & $-$ \\
$\alpha$ & $0.54(1)$          & $0.54(2)$          & $-$ \\
$b$      & $2.3(2)\,\rm MeV$  & $-1.0(4)\,\rm MeV$ & $-$ \\
$\beta$  & $2.6(1)$           & $4(1)$             & $-$ \\[0.1cm]
$n_0$    & $0.22(1)\,\rm fm^{-3}$ & $0.36(1)\,\rm fm^{-3}$ & $0.164(7)\,\rm fm^{-3}$ \\
$E_0$    & $-13.96(8)\,\rm MeV$   & $-17.29(9)\,\rm MeV$   & $-15.86(57)\,\rm MeV$ \\
$K_0$    & $223(16)\,\rm MeV$     & $184(64)\,\rm MeV$     & $230(20)\,\rm MeV$ \\
$Q_0$    & $252(390)\,\rm MeV$    & $1110(1491)\,\rm MeV$  & $300(400)\,\rm MeV$ \\
\end{tabular}
\end{ruledtabular}
\end{table}

In \cref{fig:eos-esym}, colored bands represent the uncertainties of
the many-body calculation, which include both statistical Monte Carlo
errors and the uncertainties coming from the truncation of the chiral
expansion. The latter is evaluated according to the prescription by
Epelbaum \textit{et al.}~\cite{Epelbaum:2015}.
In this work we consider the average momentum scale $p=\sqrt{3/5}\,k_F$~\cite{Lynn:2017}, 
$k_F$ being the Fermi momentum of PNM or SNM, and 
$\Lambda_b=500\,\rm MeV$~\cite{Lynn:2016}.
The difference of the two bands indicates an additional source of uncertainty due to the regularization scheme.

The EOS of SNM for the $E\mathbbm1$ potential saturates at a
slightly higher density $n=0.22(1)\,\rm fm^{-3}$ and higher energy
$E=-13.96(8)\,\rm MeV$ compared to the empirical point,
while the incompressibility $K_0=223(16)\,\rm MeV$ lies within the expected
range~\cite{Margueron:2017}.  The skewness parameter is poorly
constrained, but is consistent, for instance, with the analysis
carried out in Refs.~\cite{Margueron:2017,Etivaux:2019}, where the authors considered
terms up to $n^4$ to fit the SNM EOS.  Note that, by considering in
\cref{eq:snm} only terms up to $n^3$, and by constraining the skewness
parameter to impose the passage to zero, $n_0$, $E_0$, and $K_0$
stay within the previously identified ranges, while $Q_0$ changes to
$-145(108)\,\rm MeV$.  This value is similar to that extracted from
Skyrme parametrizations, where a similar EOS up to $n^3$ terms is used
(see, for instance, Ref.~\cite{Cai:2014kya}).

\begin{table}[t]
\centering
\caption[]{Nuclear symmetry energy and its slope (in MeV) at different densities. 
Uncertainties are at the $1\sigma$ confidence level.
Empirical values are also reported.}
\label{tab:s_l}
\begin{ruledtabular}
\begin{tabular}{lcccc}
Density & Obs. & N$^2$LO$\,E\mathbbm1$ & N$^2$LO$\,E\tau$ & Empirical\\
\midrule
$n_{\rm sat}$  & $S$           & $30(3)$  & $27(3)$ & $31.6(2.7)$~\cite{Li:2019}\\
               & $S_{\rm PNM}$ & $33(1)$  & $29(1)$ & \\
               & $L$           & $59(9)$  & $33(9)$ & $58.9(16.0)$~\cite{Li:2019}\\
               & $L_{\rm PNM}$ & $40(4)$  & $11(5)$ & \\[0.1cm]
$2n_{\rm sat}$ & $S$           & $45(5)$  & $-$     & $46-54$~\cite{Russotto:2016} \\
               & $L$           & $67(44)$ & $-$     & $-$ \\
\end{tabular}
\end{ruledtabular}
\end{table}

The new PNM EOS is consistent with earlier AFDMC results using local
chiral interactions~\cite{Lynn:2016}, where simplified wave functions
were used and the unconstrained propagation was not performed. Hence,
the PNM EOS for the $E\mathbbm1$ interaction remains stiff enough to
be compatible with astrophysical observations, while the $E\tau$
potential is too attractive at high density, resulting in negative
pressure above $0.20\,\rm fm^{-3}$.
This behavior is similar to the SNM case, where saturation is reached
at high density $[n=0.36(1)\,\rm fm^{-3}]$, with an energy below the
empirical saturation value $[-17.29(9)\,\rm MeV]$. The incompressibility
$[K_0=184(64)\,\rm MeV]$ is within the expected range, but the uncertainty
is large. The skewness parameter is also very poorly constrained.

The nuclear symmetry energy $S$ and its slope $L$ as a function of the density are defined as:
\begin{align}
    S(n)=E_{\rm PNM}(n)-E_{\rm SNM}(n), \label{eq:s}
\end{align}
\vspace{-0.6cm}
\begin{align}
    L(n)=3n\frac{\partial S(n)}{\partial n}. \label{eq:l}
\end{align}
Please note, that the symmetry energy as defined \cref{eq:s} is similar to the quadratic term in the isospin asymmetry expansion if quartic contributions are small as expected~\cite{Carbone:2014,Wellenhofer:2016}.
The AFDMC results for the symmetry energy are reported in the right panel of \cref{fig:eos-esym} for both the local chiral interactions considered in this work. The $E\mathbbm1$ potential predicts values of the symmetry energy compatible with the available constraints at both saturation~\cite{Li:2019} and twice saturation density~\cite{Russotto:2016} (see also \cref{tab:s_l}). For the $E\tau$ model, compatibility is only marginal at saturation density, and deteriorates at higher density due to the behavior of the corresponding PNM EOS above $0.20\,\rm fm^{-3}$. 

\begin{figure}[t]
\includegraphics[width=\linewidth]{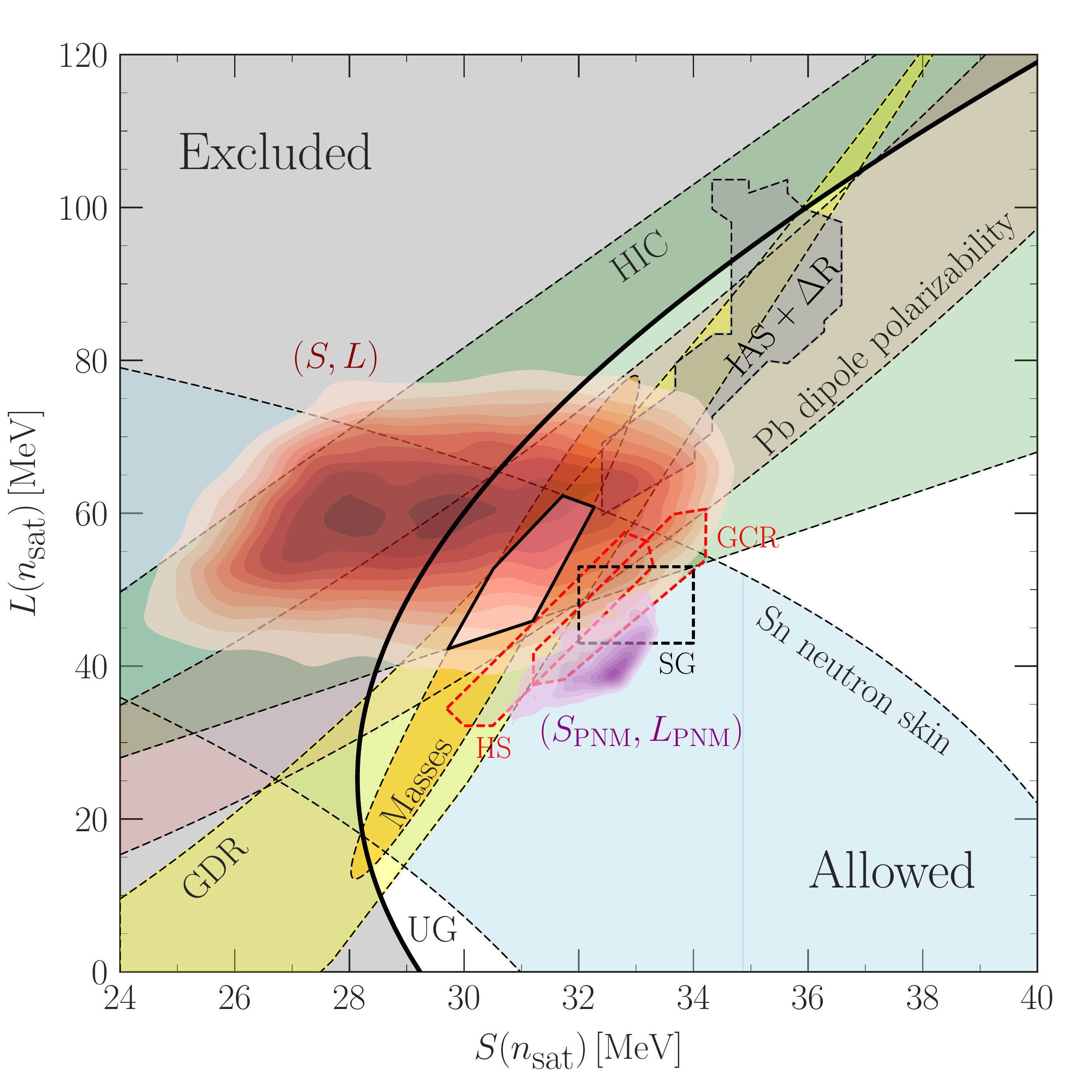}
\caption[]{$S$ vs $L$ (red) and $S_{\rm PNM}$ vs $L_{\rm PNM}$ (purple) at the empirical saturation density $n_{\rm sat}$ for the $E\mathbbm1$ potential compared to experimental constraints from nuclear masses~\cite{Kortelainen:2010hv}, the neutron-skin thicknesses of Sn isotopes~\cite{Chen:2010qx}, the dipole polarizability of \isotope[208]{Pb}~\cite{Piekarewicz:2012pp}, giant dipole resonances (GDR)~\cite{Trippa:2008gr}, isotope diffusion in heavy ion collisions (HIC)~\cite{Tsang:2008fd}, and from isobaric analog states and isovector skin (IAS+$\Delta R$)~\cite{Danielewicz:2016bgb}. The areas denoted by red-dashed lines are theoretical constraints from Ref.~\cite{Hebeler:2010jx} (HS) and Ref.~\cite{Gandolfi:2012} (GCR), while the area denoted by black-dashed lines is the inference of Ref.~\cite{Steiner:2012} (SG). The thick black line shows the unitary-gas constraint from Ref.~\cite{Tews:2017}. 
Figure adapted from Ref.~\cite{Tews:2017}.}
\label{fig:ls}
\end{figure}

In \cref{tab:s_l} we report the values of the symmetry energy and its slope at $n_{\rm sat}$ and $2n_{\rm sat}$ as obtained from \cref{eq:s,eq:l}. For the $E\mathbbm1$ interaction, $L$ is compatible with the empirical values, within the estimated uncertainties. The $E\tau$ potential, instead, predicts a too low value for $L$, as a consequence of the corresponding too soft PNM EOS. 
If the empirical saturation was achieved for the employed interactions, $n_0\equiv n_{\rm sat}$, the symmetry energy and its slope at $n_{\rm sat}$ would be completely determined by the PNM EOS: $S_{\rm PNM}=a+b-E_{\rm sat}$, with $E_{\rm sat}=-15.86\,\rm MeV$, and $L_{\rm PNM}=3(a\alpha+b\beta)$, see values in \cref{tab:s_l}. Differences between $(S,L)$ and $(S_{\rm PNM},L_{\rm PNM})$, due to extra energy and pressure contributions from the SNM EOS, are clearly shown in \cref{fig:ls}, where $L$ is plotted versus $S$ at the empirical saturation density. Shaded areas are calculated by sampling thousands of curves within the uncertainty bands of PNM and SNM, calculating $S$ and $L$ from these samples, and plotting the resulting densities. Colored bands with labels are experimental constraints as in Ref.~\cite{Tews:2017}.

\begin{figure}[t]
\includegraphics[width=\linewidth]{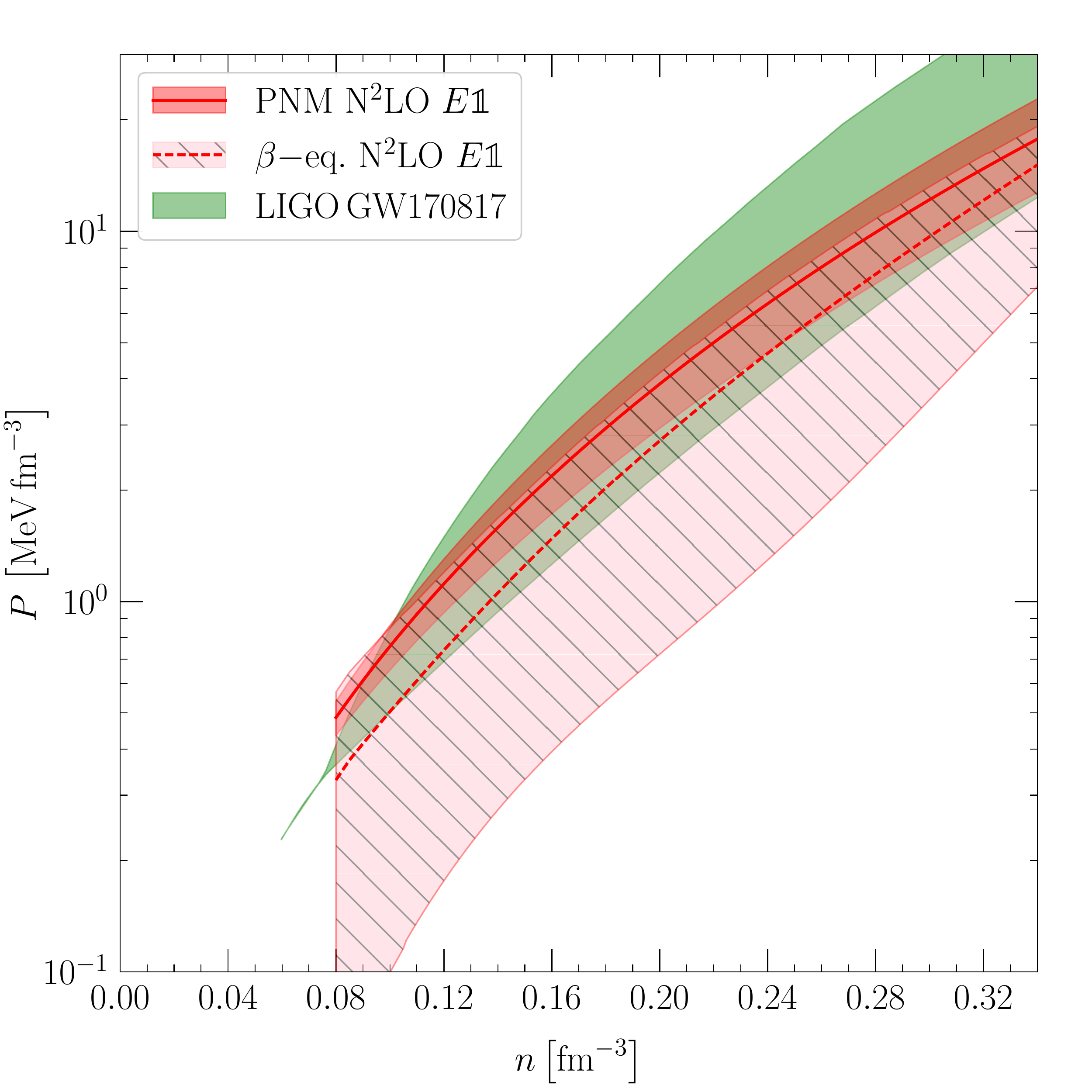}
\caption[]{Pressure of PNM (red band) and matter in $\beta$-equilibrium (red hatched band) as a function of density for the $E\mathbbm1$ interaction. The green area is the pressure extracted from the GW signal GW170817 for EOSs reproducing a $2M_\odot$ neutron star~\cite{Abbott:2018exr}.}
\label{fig:press}
\end{figure}

Finally, in \cref{fig:press}, for the $E\mathbbm1$ interaction we show the pressure as a function of the density for PNM (red solid curve) and $\beta$-equilibrated matter (red dashed curve), where the latter is obtained consistently from our results for PNM and SNM, including the uncertainty bands.
The green area is the pressure extracted by the LIGO-Virgo collaboration from the GW signal GW170817~\cite{Abbott:2018exr}. At low densities, the LIGO extraction is stitched to the SLy EOS, which is why the uncertainty band decreases. We find that our calculations lead to pressures that are compatible with but at the lower bound of the LIGO extraction.

\textit{Summary.} 
We have performed QMC calculations of symmetric nuclear
matter and the symmetry energy using realistic nuclear interactions
from chiral EFT. For our local chiral interactions at N$^2$LO, we find
saturation at $\approx1.4n_{\rm sat}$ and $\approx -14\,\rm MeV$, but our
results overlap with the empirical saturation point within uncertainties.
Our results for the symmetry energy and its density behavior agree with
previous inferences from experimental data up to $2n_{\rm sat}$, and
lead to a pressure for $\beta$-equilibrated matter in agreement with
inferences from the neutron-star merger GW170817.
We stress that the interactions employed here have been fit only to few-body systems,
i.e., nucleon-nucleon scattering, and $A=4$ and 5 nuclei, and it is quite remarkable
that the EOS we have calculated is reproducing well the constraints from GWs.

\textit{Acknowledgments.} 
We thank J.~Margueron and A.~Schwenk for insightful discussions.
The work of D.L. was supported by the U.S. Department of Energy, Office
of Science, Office of Nuclear Physics, under Contract No.~DE-SC0013617,
and by the NUCLEI SciDAC program. The work of I.T., S.G., and
J.C. was supported by the U.S. Department of Energy, Office of Science,
Office of Nuclear Physics, under Contract No.~DE-AC52-06NA25396, by
the NUCLEI SciDAC program, and by the LDRD program at LANL.  S.G. was
also supported by the DOE Early Career Research Program. Computational
resources have been provided  by the Los Alamos National Laboratory
Institutional Computing Program, which is supported by the U.S. Department
of Energy National Nuclear Security Administration under Contract
No.~89233218CNA000001, and by the National Energy Research Scientific
Computing Center (NERSC), which is supported by the U.S. Department of
Energy, Office of Science, under Contract No.~DE-AC02-05CH11231.


\begin{thebibliography}{58}%
\makeatletter
\providecommand \@ifxundefined [1]{%
 \@ifx{#1\undefined}
}%
\providecommand \@ifnum [1]{%
 \ifnum #1\expandafter \@firstoftwo
 \else \expandafter \@secondoftwo
 \fi
}%
\providecommand \@ifx [1]{%
 \ifx #1\expandafter \@firstoftwo
 \else \expandafter \@secondoftwo
 \fi
}%
\providecommand \natexlab [1]{#1}%
\providecommand \enquote  [1]{``#1''}%
\providecommand \bibnamefont  [1]{#1}%
\providecommand \bibfnamefont [1]{#1}%
\providecommand \citenamefont [1]{#1}%
\providecommand \href@noop [0]{\@secondoftwo}%
\providecommand \href [0]{\begingroup \@sanitize@url \@href}%
\providecommand \@href[1]{\@@startlink{#1}\@@href}%
\providecommand \@@href[1]{\endgroup#1\@@endlink}%
\providecommand \@sanitize@url [0]{\catcode `\\12\catcode `\$12\catcode
  `\&12\catcode `\#12\catcode `\^12\catcode `\_12\catcode `\%12\relax}%
\providecommand \@@startlink[1]{}%
\providecommand \@@endlink[0]{}%
\providecommand \url  [0]{\begingroup\@sanitize@url \@url }%
\providecommand \@url [1]{\endgroup\@href {#1}{\urlprefix }}%
\providecommand \urlprefix  [0]{URL }%
\providecommand \Eprint [0]{\href }%
\providecommand \doibase [0]{http://dx.doi.org/}%
\providecommand \selectlanguage [0]{\@gobble}%
\providecommand \bibinfo  [0]{\@secondoftwo}%
\providecommand \bibfield  [0]{\@secondoftwo}%
\providecommand \translation [1]{[#1]}%
\providecommand \BibitemOpen [0]{}%
\providecommand \bibitemStop [0]{}%
\providecommand \bibitemNoStop [0]{.\EOS\space}%
\providecommand \EOS [0]{\spacefactor3000\relax}%
\providecommand \BibitemShut  [1]{\csname bibitem#1\endcsname}%
\let\auto@bib@innerbib\@empty
\bibitem [{\citenamefont {Hebeler}\ \emph {et~al.}(2010)\citenamefont
  {Hebeler}, \citenamefont {Lattimer}, \citenamefont {Pethick},\ and\
  \citenamefont {Schwenk}}]{Hebeler:2010jx}%
  \BibitemOpen
  \bibfield  {author} {\bibinfo {author} {\bibfnamefont {K.}~\bibnamefont
  {Hebeler}}, \bibinfo {author} {\bibfnamefont {J.~M.}\ \bibnamefont
  {Lattimer}}, \bibinfo {author} {\bibfnamefont {C.~J.}\ \bibnamefont
  {Pethick}}, \ and\ \bibinfo {author} {\bibfnamefont {A.}~\bibnamefont
  {Schwenk}},\ }\href {\doibase 10.1103/PhysRevLett.105.161102} {\bibfield
  {journal} {\bibinfo  {journal} {Phys. Rev. Lett.}\ }\textbf {\bibinfo
  {volume} {105}},\ \bibinfo {pages} {161102} (\bibinfo {year}
  {2010})}\BibitemShut {NoStop}%
\bibitem [{\citenamefont {Gandolfi}\ \emph {et~al.}(2012)\citenamefont
  {Gandolfi}, \citenamefont {Carlson},\ and\ \citenamefont
  {Reddy}}]{Gandolfi:2012}%
  \BibitemOpen
  \bibfield  {author} {\bibinfo {author} {\bibfnamefont {S.}~\bibnamefont
  {Gandolfi}}, \bibinfo {author} {\bibfnamefont {J.}~\bibnamefont {Carlson}}, \
  and\ \bibinfo {author} {\bibfnamefont {S.}~\bibnamefont {Reddy}},\ }\href
  {\doibase 10.1103/PhysRevC.85.032801} {\bibfield  {journal} {\bibinfo
  {journal} {Phys. Rev. C}\ }\textbf {\bibinfo {volume} {85}},\ \bibinfo
  {pages} {032801(R)} (\bibinfo {year} {2012})}\BibitemShut {NoStop}%
\bibitem [{\citenamefont {Steiner}\ and\ \citenamefont
  {Gandolfi}(2012)}]{Steiner:2012}%
  \BibitemOpen
  \bibfield  {author} {\bibinfo {author} {\bibfnamefont {A.~W.}\ \bibnamefont
  {Steiner}}\ and\ \bibinfo {author} {\bibfnamefont {S.}~\bibnamefont
  {Gandolfi}},\ }\href {\doibase 10.1103/PhysRevLett.108.081102} {\bibfield
  {journal} {\bibinfo  {journal} {Phys. Rev. Lett.}\ }\textbf {\bibinfo
  {volume} {108}},\ \bibinfo {pages} {081102} (\bibinfo {year}
  {2012})}\BibitemShut {NoStop}%
\bibitem [{\citenamefont {Tsang}\ \emph {et~al.}(2012)\citenamefont {Tsang},
  \citenamefont {Stone}, \citenamefont {Camera}, \citenamefont {Danielewicz},
  \citenamefont {Gandolfi}, \citenamefont {Hebeler}, \citenamefont {Horowitz},
  \citenamefont {Lee}, \citenamefont {Lynch}, \citenamefont {Kohley},
  \citenamefont {Lemmon}, \citenamefont {M\"oller}, \citenamefont {Murakami},
  \citenamefont {Riordan}, \citenamefont {Roca-Maza}, \citenamefont
  {Sammarruca}, \citenamefont {Steiner}, \citenamefont {Vida\~na},\ and\
  \citenamefont {Yennello}}]{Tsang:2012}%
  \BibitemOpen
  \bibfield  {author} {\bibinfo {author} {\bibfnamefont {M.~B.}\ \bibnamefont
  {Tsang}}, \bibinfo {author} {\bibfnamefont {J.~R.}\ \bibnamefont {Stone}},
  \bibinfo {author} {\bibfnamefont {F.}~\bibnamefont {Camera}}, \bibinfo
  {author} {\bibfnamefont {P.}~\bibnamefont {Danielewicz}}, \bibinfo {author}
  {\bibfnamefont {S.}~\bibnamefont {Gandolfi}}, \bibinfo {author}
  {\bibfnamefont {K.}~\bibnamefont {Hebeler}}, \bibinfo {author} {\bibfnamefont
  {C.~J.}\ \bibnamefont {Horowitz}}, \bibinfo {author} {\bibfnamefont
  {J.}~\bibnamefont {Lee}}, \bibinfo {author} {\bibfnamefont {W.~G.}\
  \bibnamefont {Lynch}}, \bibinfo {author} {\bibfnamefont {Z.}~\bibnamefont
  {Kohley}}, \bibinfo {author} {\bibfnamefont {R.}~\bibnamefont {Lemmon}},
  \bibinfo {author} {\bibfnamefont {P.}~\bibnamefont {M\"oller}}, \bibinfo
  {author} {\bibfnamefont {T.}~\bibnamefont {Murakami}}, \bibinfo {author}
  {\bibfnamefont {S.}~\bibnamefont {Riordan}}, \bibinfo {author} {\bibfnamefont
  {X.}~\bibnamefont {Roca-Maza}}, \bibinfo {author} {\bibfnamefont
  {F.}~\bibnamefont {Sammarruca}}, \bibinfo {author} {\bibfnamefont {A.~W.}\
  \bibnamefont {Steiner}}, \bibinfo {author} {\bibfnamefont {I.}~\bibnamefont
  {Vida\~na}}, \ and\ \bibinfo {author} {\bibfnamefont {S.~J.}\ \bibnamefont
  {Yennello}},\ }\href {\doibase 10.1103/PhysRevC.86.015803} {\bibfield
  {journal} {\bibinfo  {journal} {Phys. Rev. C}\ }\textbf {\bibinfo {volume}
  {86}},\ \bibinfo {pages} {015803} (\bibinfo {year} {2012})}\BibitemShut
  {NoStop}%
\bibitem [{\citenamefont {{Lattimer}}\ and\ \citenamefont
  {{Steiner}}(2014)}]{lattimer14}%
  \BibitemOpen
  \bibfield  {author} {\bibinfo {author} {\bibfnamefont {J.~M.}\ \bibnamefont
  {{Lattimer}}}\ and\ \bibinfo {author} {\bibfnamefont {A.~W.}\ \bibnamefont
  {{Steiner}}},\ }\href {\doibase 10.1140/epja/i2014-14040-y} {\bibfield
  {journal} {\bibinfo  {journal} {Eur. Phys. J. A}\ }\textbf {\bibinfo {volume}
  {50}},\ \bibinfo {eid} {40} (\bibinfo {year} {2014})}\BibitemShut {NoStop}%
\bibitem [{\citenamefont {Lattimer}\ and\ \citenamefont
  {Lim}(2013)}]{Lattimer:2012xj}%
  \BibitemOpen
  \bibfield  {author} {\bibinfo {author} {\bibfnamefont {J.~M.}\ \bibnamefont
  {Lattimer}}\ and\ \bibinfo {author} {\bibfnamefont {Y.}~\bibnamefont {Lim}},\
  }\href {\doibase 10.1088/0004-637X/771/1/51} {\bibfield  {journal} {\bibinfo
  {journal} {Astrophys. J.}\ }\textbf {\bibinfo {volume} {771}},\ \bibinfo
  {pages} {51} (\bibinfo {year} {2013})}\BibitemShut {NoStop}%
\bibitem [{\citenamefont {Abbott}\ \emph
  {et~al.}(2017{\natexlab{a}})\citenamefont {Abbott} \emph
  {et~al.}}]{TheLIGOScientific:2017qsa}%
  \BibitemOpen
  \bibfield  {author} {\bibinfo {author} {\bibfnamefont {B.~P.}\ \bibnamefont
  {Abbott}} \emph {et~al.} (\bibinfo {collaboration} {LIGO Scientific
  Collaboration and Virgo Collaboration}),\ }\href {\doibase
  10.1103/PhysRevLett.119.161101} {\bibfield  {journal} {\bibinfo  {journal}
  {Phys. Rev. Lett.}\ }\textbf {\bibinfo {volume} {119}},\ \bibinfo {pages}
  {161101} (\bibinfo {year} {2017}{\natexlab{a}})}\BibitemShut {NoStop}%
\bibitem [{\citenamefont {Abbott}\ \emph
  {et~al.}(2017{\natexlab{b}})\citenamefont {Abbott} \emph
  {et~al.}}]{GBM:2017lvd}%
  \BibitemOpen
  \bibfield  {author} {\bibinfo {author} {\bibfnamefont {B.~P.}\ \bibnamefont
  {Abbott}} \emph {et~al.},\ }\href {\doibase 10.3847/2041-8213/aa91c9}
  {\bibfield  {journal} {\bibinfo  {journal} {Astrophys. J.}\ }\textbf
  {\bibinfo {volume} {848}},\ \bibinfo {pages} {L12} (\bibinfo {year}
  {2017}{\natexlab{b}})}\BibitemShut {NoStop}%
\bibitem [{\citenamefont {Gezerlis}\ \emph {et~al.}(2014)\citenamefont
  {Gezerlis}, \citenamefont {Tews}, \citenamefont {Epelbaum}, \citenamefont
  {Freunek}, \citenamefont {Gandolfi}, \citenamefont {Hebeler}, \citenamefont
  {Nogga},\ and\ \citenamefont {Schwenk}}]{Gezerlis:2014}%
  \BibitemOpen
  \bibfield  {author} {\bibinfo {author} {\bibfnamefont {A.}~\bibnamefont
  {Gezerlis}}, \bibinfo {author} {\bibfnamefont {I.}~\bibnamefont {Tews}},
  \bibinfo {author} {\bibfnamefont {E.}~\bibnamefont {Epelbaum}}, \bibinfo
  {author} {\bibfnamefont {M.}~\bibnamefont {Freunek}}, \bibinfo {author}
  {\bibfnamefont {S.}~\bibnamefont {Gandolfi}}, \bibinfo {author}
  {\bibfnamefont {K.}~\bibnamefont {Hebeler}}, \bibinfo {author} {\bibfnamefont
  {A.}~\bibnamefont {Nogga}}, \ and\ \bibinfo {author} {\bibfnamefont
  {A.}~\bibnamefont {Schwenk}},\ }\href {\doibase 10.1103/PhysRevC.90.054323}
  {\bibfield  {journal} {\bibinfo  {journal} {Phys. Rev. C}\ }\textbf {\bibinfo
  {volume} {90}},\ \bibinfo {pages} {054323} (\bibinfo {year}
  {2014})}\BibitemShut {NoStop}%
\bibitem [{\citenamefont {Lynn}\ \emph {et~al.}(2016)\citenamefont {Lynn},
  \citenamefont {Tews}, \citenamefont {Carlson}, \citenamefont {Gandolfi},
  \citenamefont {Gezerlis}, \citenamefont {Schmidt},\ and\ \citenamefont
  {Schwenk}}]{Lynn:2016}%
  \BibitemOpen
  \bibfield  {author} {\bibinfo {author} {\bibfnamefont {J.~E.}\ \bibnamefont
  {Lynn}}, \bibinfo {author} {\bibfnamefont {I.}~\bibnamefont {Tews}}, \bibinfo
  {author} {\bibfnamefont {J.}~\bibnamefont {Carlson}}, \bibinfo {author}
  {\bibfnamefont {S.}~\bibnamefont {Gandolfi}}, \bibinfo {author}
  {\bibfnamefont {A.}~\bibnamefont {Gezerlis}}, \bibinfo {author}
  {\bibfnamefont {K.~E.}\ \bibnamefont {Schmidt}}, \ and\ \bibinfo {author}
  {\bibfnamefont {A.}~\bibnamefont {Schwenk}},\ }\href {\doibase
  10.1103/PhysRevLett.116.062501} {\bibfield  {journal} {\bibinfo  {journal}
  {Phys. Rev. Lett.}\ }\textbf {\bibinfo {volume} {116}},\ \bibinfo {pages}
  {062501} (\bibinfo {year} {2016})}\BibitemShut {NoStop}%
\bibitem [{\citenamefont {Akmal}\ \emph {et~al.}(1998)\citenamefont {Akmal},
  \citenamefont {Pandharipande},\ and\ \citenamefont {Ravenhall}}]{Akmal:1998}%
  \BibitemOpen
  \bibfield  {author} {\bibinfo {author} {\bibfnamefont {A.}~\bibnamefont
  {Akmal}}, \bibinfo {author} {\bibfnamefont {V.~R.}\ \bibnamefont
  {Pandharipande}}, \ and\ \bibinfo {author} {\bibfnamefont {D.~G.}\
  \bibnamefont {Ravenhall}},\ }\href {\doibase 10.1103/PhysRevC.58.1804}
  {\bibfield  {journal} {\bibinfo  {journal} {Phys. Rev. C}\ }\textbf {\bibinfo
  {volume} {58}},\ \bibinfo {pages} {1804} (\bibinfo {year}
  {1998})}\BibitemShut {NoStop}%
\bibitem [{\citenamefont {Hebeler}\ \emph {et~al.}(2011)\citenamefont
  {Hebeler}, \citenamefont {Bogner}, \citenamefont {Furnstahl}, \citenamefont
  {Nogga},\ and\ \citenamefont {Schwenk}}]{Hebeler:2010xb}%
  \BibitemOpen
  \bibfield  {author} {\bibinfo {author} {\bibfnamefont {K.}~\bibnamefont
  {Hebeler}}, \bibinfo {author} {\bibfnamefont {S.~K.}\ \bibnamefont {Bogner}},
  \bibinfo {author} {\bibfnamefont {R.~J.}\ \bibnamefont {Furnstahl}}, \bibinfo
  {author} {\bibfnamefont {A.}~\bibnamefont {Nogga}}, \ and\ \bibinfo {author}
  {\bibfnamefont {A.}~\bibnamefont {Schwenk}},\ }\href {\doibase
  10.1103/PhysRevC.83.031301} {\bibfield  {journal} {\bibinfo  {journal} {Phys.
  Rev. C}\ }\textbf {\bibinfo {volume} {C83}},\ \bibinfo {pages} {031301(R)}
  (\bibinfo {year} {2011})}\BibitemShut {NoStop}%
\bibitem [{\citenamefont {Hagen}\ \emph {et~al.}(2014)\citenamefont {Hagen},
  \citenamefont {Papenbrock}, \citenamefont {Ekstr\"om}, \citenamefont {Wendt},
  \citenamefont {Baardsen}, \citenamefont {Gandolfi}, \citenamefont
  {Hjorth-Jensen},\ and\ \citenamefont {Horowitz}}]{Hagen:2014}%
  \BibitemOpen
  \bibfield  {author} {\bibinfo {author} {\bibfnamefont {G.}~\bibnamefont
  {Hagen}}, \bibinfo {author} {\bibfnamefont {T.}~\bibnamefont {Papenbrock}},
  \bibinfo {author} {\bibfnamefont {A.}~\bibnamefont {Ekstr\"om}}, \bibinfo
  {author} {\bibfnamefont {K.~A.}\ \bibnamefont {Wendt}}, \bibinfo {author}
  {\bibfnamefont {G.}~\bibnamefont {Baardsen}}, \bibinfo {author}
  {\bibfnamefont {S.}~\bibnamefont {Gandolfi}}, \bibinfo {author}
  {\bibfnamefont {M.}~\bibnamefont {Hjorth-Jensen}}, \ and\ \bibinfo {author}
  {\bibfnamefont {C.~J.}\ \bibnamefont {Horowitz}},\ }\href {\doibase
  10.1103/PhysRevC.89.014319} {\bibfield  {journal} {\bibinfo  {journal} {Phys.
  Rev. C}\ }\textbf {\bibinfo {volume} {89}},\ \bibinfo {pages} {014319}
  (\bibinfo {year} {2014})}\BibitemShut {NoStop}%
\bibitem [{\citenamefont {Ekstr\"om}\ \emph {et~al.}(2015)\citenamefont
  {Ekstr\"om}, \citenamefont {Jansen}, \citenamefont {Wendt}, \citenamefont
  {Hagen}, \citenamefont {Papenbrock}, \citenamefont {Carlsson}, \citenamefont
  {Forss\'en}, \citenamefont {Hjorth-Jensen}, \citenamefont {Navr\'atil},\ and\
  \citenamefont {Nazarewicz}}]{Ekstrom:2015}%
  \BibitemOpen
  \bibfield  {author} {\bibinfo {author} {\bibfnamefont {A.}~\bibnamefont
  {Ekstr\"om}}, \bibinfo {author} {\bibfnamefont {G.~R.}\ \bibnamefont
  {Jansen}}, \bibinfo {author} {\bibfnamefont {K.~A.}\ \bibnamefont {Wendt}},
  \bibinfo {author} {\bibfnamefont {G.}~\bibnamefont {Hagen}}, \bibinfo
  {author} {\bibfnamefont {T.}~\bibnamefont {Papenbrock}}, \bibinfo {author}
  {\bibfnamefont {B.~D.}\ \bibnamefont {Carlsson}}, \bibinfo {author}
  {\bibfnamefont {C.}~\bibnamefont {Forss\'en}}, \bibinfo {author}
  {\bibfnamefont {M.}~\bibnamefont {Hjorth-Jensen}}, \bibinfo {author}
  {\bibfnamefont {P.}~\bibnamefont {Navr\'atil}}, \ and\ \bibinfo {author}
  {\bibfnamefont {W.}~\bibnamefont {Nazarewicz}},\ }\href {\doibase
  10.1103/PhysRevC.91.051301} {\bibfield  {journal} {\bibinfo  {journal} {Phys.
  Rev. C}\ }\textbf {\bibinfo {volume} {91}},\ \bibinfo {pages} {051301(R)}
  (\bibinfo {year} {2015})}\BibitemShut {NoStop}%
\bibitem [{\citenamefont {Holt}\ and\ \citenamefont
  {Kaiser}(2017)}]{Holt:2016pjb}%
  \BibitemOpen
  \bibfield  {author} {\bibinfo {author} {\bibfnamefont {J.~W.}\ \bibnamefont
  {Holt}}\ and\ \bibinfo {author} {\bibfnamefont {N.}~\bibnamefont {Kaiser}},\
  }\href {\doibase 10.1103/PhysRevC.95.034326} {\bibfield  {journal} {\bibinfo
  {journal} {Phys. Rev. C}\ }\textbf {\bibinfo {volume} {95}},\ \bibinfo
  {pages} {034326} (\bibinfo {year} {2017})}\BibitemShut {NoStop}%
\bibitem [{\citenamefont {Drischler}\ \emph {et~al.}(2019)\citenamefont
  {Drischler}, \citenamefont {Hebeler},\ and\ \citenamefont
  {Schwenk}}]{Drischler:2017wtt}%
  \BibitemOpen
  \bibfield  {author} {\bibinfo {author} {\bibfnamefont {C.}~\bibnamefont
  {Drischler}}, \bibinfo {author} {\bibfnamefont {K.}~\bibnamefont {Hebeler}},
  \ and\ \bibinfo {author} {\bibfnamefont {A.}~\bibnamefont {Schwenk}},\ }\href
  {\doibase 10.1103/PhysRevLett.122.042501} {\bibfield  {journal} {\bibinfo
  {journal} {Phys. Rev. Lett.}\ }\textbf {\bibinfo {volume} {122}},\ \bibinfo
  {pages} {042501} (\bibinfo {year} {2019})}\BibitemShut {NoStop}%
\bibitem [{\citenamefont {Carbone}\ and\ \citenamefont
  {Schwenk}(2019)}]{Carbone:2019}%
  \BibitemOpen
  \bibfield  {author} {\bibinfo {author} {\bibfnamefont {A.}~\bibnamefont
  {Carbone}}\ and\ \bibinfo {author} {\bibfnamefont {A.}~\bibnamefont
  {Schwenk}},\ }\href {\doibase 10.1103/PhysRevC.100.025805} {\bibfield
  {journal} {\bibinfo  {journal} {Phys. Rev. C}\ }\textbf {\bibinfo {volume}
  {100}},\ \bibinfo {pages} {025805} (\bibinfo {year} {2019})}\BibitemShut
  {NoStop}%
\bibitem [{\citenamefont {Carlson}\ \emph {et~al.}(2015)\citenamefont
  {Carlson}, \citenamefont {Gandolfi}, \citenamefont {Pederiva}, \citenamefont
  {Pieper}, \citenamefont {Schiavilla}, \citenamefont {Schmidt},\ and\
  \citenamefont {Wiringa}}]{Carlson:2015}%
  \BibitemOpen
  \bibfield  {author} {\bibinfo {author} {\bibfnamefont {J.}~\bibnamefont
  {Carlson}}, \bibinfo {author} {\bibfnamefont {S.}~\bibnamefont {Gandolfi}},
  \bibinfo {author} {\bibfnamefont {F.}~\bibnamefont {Pederiva}}, \bibinfo
  {author} {\bibfnamefont {S.~C.}\ \bibnamefont {Pieper}}, \bibinfo {author}
  {\bibfnamefont {R.}~\bibnamefont {Schiavilla}}, \bibinfo {author}
  {\bibfnamefont {K.~E.}\ \bibnamefont {Schmidt}}, \ and\ \bibinfo {author}
  {\bibfnamefont {R.~B.}\ \bibnamefont {Wiringa}},\ }\href {\doibase
  10.1103/RevModPhys.87.1067} {\bibfield  {journal} {\bibinfo  {journal} {Rev.
  Mod. Phys.}\ }\textbf {\bibinfo {volume} {87}},\ \bibinfo {pages} {1067}
  (\bibinfo {year} {2015})}\BibitemShut {NoStop}%
\bibitem [{\citenamefont {Gezerlis}\ \emph {et~al.}(2013)\citenamefont
  {Gezerlis}, \citenamefont {Tews}, \citenamefont {Epelbaum}, \citenamefont
  {Gandolfi}, \citenamefont {Hebeler}, \citenamefont {Nogga},\ and\
  \citenamefont {Schwenk}}]{Gezerlis:2013}%
  \BibitemOpen
  \bibfield  {author} {\bibinfo {author} {\bibfnamefont {A.}~\bibnamefont
  {Gezerlis}}, \bibinfo {author} {\bibfnamefont {I.}~\bibnamefont {Tews}},
  \bibinfo {author} {\bibfnamefont {E.}~\bibnamefont {Epelbaum}}, \bibinfo
  {author} {\bibfnamefont {S.}~\bibnamefont {Gandolfi}}, \bibinfo {author}
  {\bibfnamefont {K.}~\bibnamefont {Hebeler}}, \bibinfo {author} {\bibfnamefont
  {A.}~\bibnamefont {Nogga}}, \ and\ \bibinfo {author} {\bibfnamefont
  {A.}~\bibnamefont {Schwenk}},\ }\href {\doibase
  10.1103/PhysRevLett.111.032501} {\bibfield  {journal} {\bibinfo  {journal}
  {Phys. Rev. Lett.}\ }\textbf {\bibinfo {volume} {111}},\ \bibinfo {pages}
  {032501} (\bibinfo {year} {2013})}\BibitemShut {NoStop}%
\bibitem [{\citenamefont {Lynn}\ \emph {et~al.}(2017)\citenamefont {Lynn},
  \citenamefont {Tews}, \citenamefont {Carlson}, \citenamefont {Gandolfi},
  \citenamefont {Gezerlis}, \citenamefont {Schmidt},\ and\ \citenamefont
  {Schwenk}}]{Lynn:2017}%
  \BibitemOpen
  \bibfield  {author} {\bibinfo {author} {\bibfnamefont {J.~E.}\ \bibnamefont
  {Lynn}}, \bibinfo {author} {\bibfnamefont {I.}~\bibnamefont {Tews}}, \bibinfo
  {author} {\bibfnamefont {J.}~\bibnamefont {Carlson}}, \bibinfo {author}
  {\bibfnamefont {S.}~\bibnamefont {Gandolfi}}, \bibinfo {author}
  {\bibfnamefont {A.}~\bibnamefont {Gezerlis}}, \bibinfo {author}
  {\bibfnamefont {K.~E.}\ \bibnamefont {Schmidt}}, \ and\ \bibinfo {author}
  {\bibfnamefont {A.}~\bibnamefont {Schwenk}},\ }\href {\doibase
  10.1103/PhysRevC.96.054007} {\bibfield  {journal} {\bibinfo  {journal} {Phys.
  Rev. C}\ }\textbf {\bibinfo {volume} {96}},\ \bibinfo {pages} {054007}
  (\bibinfo {year} {2017})}\BibitemShut {NoStop}%
\bibitem [{\citenamefont {Lonardoni}\ \emph
  {et~al.}(2018{\natexlab{a}})\citenamefont {Lonardoni}, \citenamefont
  {Carlson}, \citenamefont {Gandolfi}, \citenamefont {Lynn}, \citenamefont
  {Schmidt}, \citenamefont {Schwenk},\ and\ \citenamefont
  {Wang}}]{Lonardoni:2018prl}%
  \BibitemOpen
  \bibfield  {author} {\bibinfo {author} {\bibfnamefont {D.}~\bibnamefont
  {Lonardoni}}, \bibinfo {author} {\bibfnamefont {J.}~\bibnamefont {Carlson}},
  \bibinfo {author} {\bibfnamefont {S.}~\bibnamefont {Gandolfi}}, \bibinfo
  {author} {\bibfnamefont {J.~E.}\ \bibnamefont {Lynn}}, \bibinfo {author}
  {\bibfnamefont {K.~E.}\ \bibnamefont {Schmidt}}, \bibinfo {author}
  {\bibfnamefont {A.}~\bibnamefont {Schwenk}}, \ and\ \bibinfo {author}
  {\bibfnamefont {X.~B.}\ \bibnamefont {Wang}},\ }\href {\doibase
  10.1103/PhysRevLett.120.122502} {\bibfield  {journal} {\bibinfo  {journal}
  {Phys. Rev. Lett.}\ }\textbf {\bibinfo {volume} {120}},\ \bibinfo {pages}
  {122502} (\bibinfo {year} {2018}{\natexlab{a}})}\BibitemShut {NoStop}%
\bibitem [{\citenamefont {Lonardoni}\ \emph
  {et~al.}(2018{\natexlab{b}})\citenamefont {Lonardoni}, \citenamefont
  {Gandolfi}, \citenamefont {Lynn}, \citenamefont {Petrie}, \citenamefont
  {Carlson}, \citenamefont {Schmidt},\ and\ \citenamefont
  {Schwenk}}]{Lonardoni:2018prc}%
  \BibitemOpen
  \bibfield  {author} {\bibinfo {author} {\bibfnamefont {D.}~\bibnamefont
  {Lonardoni}}, \bibinfo {author} {\bibfnamefont {S.}~\bibnamefont {Gandolfi}},
  \bibinfo {author} {\bibfnamefont {J.~E.}\ \bibnamefont {Lynn}}, \bibinfo
  {author} {\bibfnamefont {C.}~\bibnamefont {Petrie}}, \bibinfo {author}
  {\bibfnamefont {J.}~\bibnamefont {Carlson}}, \bibinfo {author} {\bibfnamefont
  {K.~E.}\ \bibnamefont {Schmidt}}, \ and\ \bibinfo {author} {\bibfnamefont
  {A.}~\bibnamefont {Schwenk}},\ }\href {\doibase 10.1103/PhysRevC.97.044318}
  {\bibfield  {journal} {\bibinfo  {journal} {Phys. Rev. C}\ }\textbf {\bibinfo
  {volume} {97}},\ \bibinfo {pages} {044318} (\bibinfo {year}
  {2018}{\natexlab{b}})}\BibitemShut {NoStop}%
\bibitem [{\citenamefont {Zhao}\ and\ \citenamefont
  {Gandolfi}(2016)}]{Gandolfi:2016}%
  \BibitemOpen
  \bibfield  {author} {\bibinfo {author} {\bibfnamefont {P.~W.}\ \bibnamefont
  {Zhao}}\ and\ \bibinfo {author} {\bibfnamefont {S.}~\bibnamefont
  {Gandolfi}},\ }\href {\doibase 10.1103/PhysRevC.94.041302} {\bibfield
  {journal} {\bibinfo  {journal} {Phys. Rev. C}\ }\textbf {\bibinfo {volume}
  {94}},\ \bibinfo {pages} {041302(R)} (\bibinfo {year} {2016})}\BibitemShut
  {NoStop}%
\bibitem [{\citenamefont {Gandolfi}\ \emph {et~al.}(2017)\citenamefont
  {Gandolfi}, \citenamefont {Hammer}, \citenamefont {Klos}, \citenamefont
  {Lynn},\ and\ \citenamefont {Schwenk}}]{Gandolfi:2017}%
  \BibitemOpen
  \bibfield  {author} {\bibinfo {author} {\bibfnamefont {S.}~\bibnamefont
  {Gandolfi}}, \bibinfo {author} {\bibfnamefont {H.-W.}\ \bibnamefont
  {Hammer}}, \bibinfo {author} {\bibfnamefont {P.}~\bibnamefont {Klos}},
  \bibinfo {author} {\bibfnamefont {J.~E.}\ \bibnamefont {Lynn}}, \ and\
  \bibinfo {author} {\bibfnamefont {A.}~\bibnamefont {Schwenk}},\ }\href
  {\doibase 10.1103/PhysRevLett.118.232501} {\bibfield  {journal} {\bibinfo
  {journal} {Phys. Rev. Lett.}\ }\textbf {\bibinfo {volume} {118}},\ \bibinfo
  {pages} {232501} (\bibinfo {year} {2017})}\BibitemShut {NoStop}%
\bibitem [{\citenamefont {Tews}\ \emph {et~al.}(2016)\citenamefont {Tews},
  \citenamefont {Gandolfi}, \citenamefont {Gezerlis},\ and\ \citenamefont
  {Schwenk}}]{Tews:2016}%
  \BibitemOpen
  \bibfield  {author} {\bibinfo {author} {\bibfnamefont {I.}~\bibnamefont
  {Tews}}, \bibinfo {author} {\bibfnamefont {S.}~\bibnamefont {Gandolfi}},
  \bibinfo {author} {\bibfnamefont {A.}~\bibnamefont {Gezerlis}}, \ and\
  \bibinfo {author} {\bibfnamefont {A.}~\bibnamefont {Schwenk}},\ }\href
  {\doibase 10.1103/PhysRevC.93.024305} {\bibfield  {journal} {\bibinfo
  {journal} {Phys. Rev. C}\ }\textbf {\bibinfo {volume} {93}},\ \bibinfo
  {pages} {024305} (\bibinfo {year} {2016})}\BibitemShut {NoStop}%
\bibitem [{\citenamefont {Buraczynski}\ and\ \citenamefont
  {Gezerlis}(2016)}]{Buraczynski:2016}%
  \BibitemOpen
  \bibfield  {author} {\bibinfo {author} {\bibfnamefont {M.}~\bibnamefont
  {Buraczynski}}\ and\ \bibinfo {author} {\bibfnamefont {A.}~\bibnamefont
  {Gezerlis}},\ }\href {\doibase 10.1103/PhysRevLett.116.152501} {\bibfield
  {journal} {\bibinfo  {journal} {Phys. Rev. Lett.}\ }\textbf {\bibinfo
  {volume} {116}},\ \bibinfo {pages} {152501} (\bibinfo {year}
  {2016})}\BibitemShut {NoStop}%
\bibitem [{\citenamefont {Tews}\ \emph
  {et~al.}(2018{\natexlab{a}})\citenamefont {Tews}, \citenamefont {Carlson},
  \citenamefont {Gandolfi},\ and\ \citenamefont {Reddy}}]{Tews:2018apj}%
  \BibitemOpen
  \bibfield  {author} {\bibinfo {author} {\bibfnamefont {I.}~\bibnamefont
  {Tews}}, \bibinfo {author} {\bibfnamefont {J.}~\bibnamefont {Carlson}},
  \bibinfo {author} {\bibfnamefont {S.}~\bibnamefont {Gandolfi}}, \ and\
  \bibinfo {author} {\bibfnamefont {S.}~\bibnamefont {Reddy}},\ }\href
  {\doibase 10.3847/1538-4357/aac267} {\bibfield  {journal} {\bibinfo
  {journal} {Astrophys. J.}\ }\textbf {\bibinfo {volume} {860}},\ \bibinfo
  {pages} {149} (\bibinfo {year} {2018}{\natexlab{a}})}\BibitemShut {NoStop}%
\bibitem [{\citenamefont {Tews}\ \emph
  {et~al.}(2018{\natexlab{b}})\citenamefont {Tews}, \citenamefont {Margueron},\
  and\ \citenamefont {Reddy}}]{Tews:2018}%
  \BibitemOpen
  \bibfield  {author} {\bibinfo {author} {\bibfnamefont {I.}~\bibnamefont
  {Tews}}, \bibinfo {author} {\bibfnamefont {J.}~\bibnamefont {Margueron}}, \
  and\ \bibinfo {author} {\bibfnamefont {S.}~\bibnamefont {Reddy}},\ }\href
  {\doibase 10.1103/PhysRevC.98.045804} {\bibfield  {journal} {\bibinfo
  {journal} {Phys. Rev. C}\ }\textbf {\bibinfo {volume} {98}},\ \bibinfo
  {pages} {045804} (\bibinfo {year} {2018}{\natexlab{b}})}\BibitemShut
  {NoStop}%
\bibitem [{\citenamefont {Tews}\ \emph {et~al.}(2019)\citenamefont {Tews},
  \citenamefont {Margueron},\ and\ \citenamefont {Reddy}}]{Tews:2019}%
  \BibitemOpen
  \bibfield  {author} {\bibinfo {author} {\bibfnamefont {I.}~\bibnamefont
  {Tews}}, \bibinfo {author} {\bibfnamefont {J.}~\bibnamefont {Margueron}}, \
  and\ \bibinfo {author} {\bibfnamefont {S.}~\bibnamefont {Reddy}},\ }\href
  {\doibase 10.1140/epja/i2019-12774-6} {\bibfield  {journal} {\bibinfo
  {journal} {Eur. Phys. J. A}\ }\textbf {\bibinfo {volume} {55}},\ \bibinfo
  {pages} {97} (\bibinfo {year} {2019})}\BibitemShut {NoStop}%
\bibitem [{\citenamefont {Gandolfi}\ \emph {et~al.}(2014)\citenamefont
  {Gandolfi}, \citenamefont {Lovato}, \citenamefont {Carlson},\ and\
  \citenamefont {Schmidt}}]{Gandolfi:2014}%
  \BibitemOpen
  \bibfield  {author} {\bibinfo {author} {\bibfnamefont {S.}~\bibnamefont
  {Gandolfi}}, \bibinfo {author} {\bibfnamefont {A.}~\bibnamefont {Lovato}},
  \bibinfo {author} {\bibfnamefont {J.}~\bibnamefont {Carlson}}, \ and\
  \bibinfo {author} {\bibfnamefont {K.~E.}\ \bibnamefont {Schmidt}},\ }\href
  {\doibase 10.1103/PhysRevC.90.061306} {\bibfield  {journal} {\bibinfo
  {journal} {Phys. Rev. C}\ }\textbf {\bibinfo {volume} {90}},\ \bibinfo
  {pages} {061306(R)} (\bibinfo {year} {2014})}\BibitemShut {NoStop}%
\bibitem [{\citenamefont {Gandolfi}\ \emph {et~al.}(2009)\citenamefont
  {Gandolfi}, \citenamefont {Illarionov}, \citenamefont {Schmidt},
  \citenamefont {Pederiva},\ and\ \citenamefont {Fantoni}}]{Gandolfi:2009}%
  \BibitemOpen
  \bibfield  {author} {\bibinfo {author} {\bibfnamefont {S.}~\bibnamefont
  {Gandolfi}}, \bibinfo {author} {\bibfnamefont {A.~Y.}\ \bibnamefont
  {Illarionov}}, \bibinfo {author} {\bibfnamefont {K.~E.}\ \bibnamefont
  {Schmidt}}, \bibinfo {author} {\bibfnamefont {F.}~\bibnamefont {Pederiva}}, \
  and\ \bibinfo {author} {\bibfnamefont {S.}~\bibnamefont {Fantoni}},\ }\href
  {\doibase 10.1103/PhysRevC.79.054005} {\bibfield  {journal} {\bibinfo
  {journal} {Phys. Rev. C}\ }\textbf {\bibinfo {volume} {79}},\ \bibinfo
  {pages} {054005} (\bibinfo {year} {2009})}\BibitemShut {NoStop}%
\bibitem [{\citenamefont {Sarsa}\ \emph {et~al.}(2003)\citenamefont {Sarsa},
  \citenamefont {Fantoni}, \citenamefont {Schmidt},\ and\ \citenamefont
  {Pederiva}}]{Sarsa:2003}%
  \BibitemOpen
  \bibfield  {author} {\bibinfo {author} {\bibfnamefont {A.}~\bibnamefont
  {Sarsa}}, \bibinfo {author} {\bibfnamefont {S.}~\bibnamefont {Fantoni}},
  \bibinfo {author} {\bibfnamefont {K.~E.}\ \bibnamefont {Schmidt}}, \ and\
  \bibinfo {author} {\bibfnamefont {F.}~\bibnamefont {Pederiva}},\ }\href
  {\doibase 10.1103/PhysRevC.68.024308} {\bibfield  {journal} {\bibinfo
  {journal} {Phys. Rev. C}\ }\textbf {\bibinfo {volume} {68}},\ \bibinfo
  {pages} {024308} (\bibinfo {year} {2003})}\BibitemShut {NoStop}%
\bibitem [{\citenamefont {Sorella}(2001)}]{Sorella:2001}%
  \BibitemOpen
  \bibfield  {author} {\bibinfo {author} {\bibfnamefont {S.}~\bibnamefont
  {Sorella}},\ }\href {\doibase 10.1103/PhysRevB.64.024512} {\bibfield
  {journal} {\bibinfo  {journal} {Phys. Rev. B}\ }\textbf {\bibinfo {volume}
  {64}},\ \bibinfo {pages} {024512} (\bibinfo {year} {2001})}\BibitemShut
  {NoStop}%
\bibitem [{\citenamefont {Schmidt}\ and\ \citenamefont
  {Fantoni}(1999)}]{Schmidt:1999}%
  \BibitemOpen
  \bibfield  {author} {\bibinfo {author} {\bibfnamefont {K.~E.}\ \bibnamefont
  {Schmidt}}\ and\ \bibinfo {author} {\bibfnamefont {S.}~\bibnamefont
  {Fantoni}},\ }\href {\doibase 10.1016/S0370-2693(98)01522-6} {\bibfield
  {journal} {\bibinfo  {journal} {Phys. Lett. B}\ }\textbf {\bibinfo {volume}
  {446}},\ \bibinfo {pages} {99} (\bibinfo {year} {1999})}\BibitemShut
  {NoStop}%
\bibitem [{\citenamefont {Zhang}\ and\ \citenamefont
  {Krakauer}(2003)}]{Zhang:2003}%
  \BibitemOpen
  \bibfield  {author} {\bibinfo {author} {\bibfnamefont {S.}~\bibnamefont
  {Zhang}}\ and\ \bibinfo {author} {\bibfnamefont {H.}~\bibnamefont
  {Krakauer}},\ }\href {\doibase 10.1103/PhysRevLett.90.136401} {\bibfield
  {journal} {\bibinfo  {journal} {Phys. Rev. Lett.}\ }\textbf {\bibinfo
  {volume} {90}},\ \bibinfo {pages} {136401} (\bibinfo {year}
  {2003})}\BibitemShut {NoStop}%
\bibitem [{\citenamefont {Gandolfi}\ \emph {et~al.}(2015)\citenamefont
  {Gandolfi}, \citenamefont {Gezerlis},\ and\ \citenamefont
  {Carlson}}]{Gandolfi:2015}%
  \BibitemOpen
  \bibfield  {author} {\bibinfo {author} {\bibfnamefont {S.}~\bibnamefont
  {Gandolfi}}, \bibinfo {author} {\bibfnamefont {A.}~\bibnamefont {Gezerlis}},
  \ and\ \bibinfo {author} {\bibfnamefont {J.}~\bibnamefont {Carlson}},\ }\href
  {\doibase 10.1146/annurev-nucl-102014-021957} {\bibfield  {journal} {\bibinfo
   {journal} {Annu. Rev. Nucl. Part. Sci.}\ }\textbf {\bibinfo {volume} {65}},\
  \bibinfo {pages} {303} (\bibinfo {year} {2015})}\BibitemShut {NoStop}%
\bibitem [{\citenamefont {Piarulli}\ \emph {et~al.}(2020)\citenamefont
  {Piarulli}, \citenamefont {Bombaci}, \citenamefont {Logoteta}, \citenamefont
  {Lovato},\ and\ \citenamefont {Wiringa}}]{Piarulli:2019}%
  \BibitemOpen
  \bibfield  {author} {\bibinfo {author} {\bibfnamefont {M.}~\bibnamefont
  {Piarulli}}, \bibinfo {author} {\bibfnamefont {I.}~\bibnamefont {Bombaci}},
  \bibinfo {author} {\bibfnamefont {D.}~\bibnamefont {Logoteta}}, \bibinfo
  {author} {\bibfnamefont {A.}~\bibnamefont {Lovato}}, \ and\ \bibinfo {author}
  {\bibfnamefont {R.~B.}\ \bibnamefont {Wiringa}},\ }\href {\doibase
  10.1103/PhysRevC.101.045801} {\bibfield  {journal} {\bibinfo  {journal}
  {Phys. Rev. C}\ }\textbf {\bibinfo {volume} {101}},\ \bibinfo {pages}
  {045801} (\bibinfo {year} {2020})}\BibitemShut {NoStop}%
\bibitem [{\citenamefont {Brualla}\ \emph {et~al.}(2003)\citenamefont
  {Brualla}, \citenamefont {Fantoni}, \citenamefont {Sarsa}, \citenamefont
  {Schmidt},\ and\ \citenamefont {Vitiello}}]{Brualla:2003}%
  \BibitemOpen
  \bibfield  {author} {\bibinfo {author} {\bibfnamefont {L.}~\bibnamefont
  {Brualla}}, \bibinfo {author} {\bibfnamefont {S.}~\bibnamefont {Fantoni}},
  \bibinfo {author} {\bibfnamefont {A.}~\bibnamefont {Sarsa}}, \bibinfo
  {author} {\bibfnamefont {K.~E.}\ \bibnamefont {Schmidt}}, \ and\ \bibinfo
  {author} {\bibfnamefont {S.~A.}\ \bibnamefont {Vitiello}},\ }\href {\doibase
  10.1103/PhysRevC.67.065806} {\bibfield  {journal} {\bibinfo  {journal} {Phys.
  Rev. C}\ }\textbf {\bibinfo {volume} {67}},\ \bibinfo {pages} {065806}
  (\bibinfo {year} {2003})}\BibitemShut {NoStop}%
\bibitem [{\citenamefont {Schmidt}\ \emph {et~al.}(1981)\citenamefont
  {Schmidt}, \citenamefont {Lee}, \citenamefont {Kalos},\ and\ \citenamefont
  {Chester}}]{Schmidt:1981}%
  \BibitemOpen
  \bibfield  {author} {\bibinfo {author} {\bibfnamefont {K.~E.}\ \bibnamefont
  {Schmidt}}, \bibinfo {author} {\bibfnamefont {M.~A.}\ \bibnamefont {Lee}},
  \bibinfo {author} {\bibfnamefont {M.~H.}\ \bibnamefont {Kalos}}, \ and\
  \bibinfo {author} {\bibfnamefont {G.~V.}\ \bibnamefont {Chester}},\ }\href
  {\doibase 10.1103/PhysRevLett.47.807} {\bibfield  {journal} {\bibinfo
  {journal} {Phys. Rev. Lett.}\ }\textbf {\bibinfo {volume} {47}},\ \bibinfo
  {pages} {807} (\bibinfo {year} {1981})}\BibitemShut {NoStop}%
\bibitem [{\citenamefont {Zhang}\ and\ \citenamefont
  {Chen}(2015)}]{Zhang:2015}%
  \BibitemOpen
  \bibfield  {author} {\bibinfo {author} {\bibfnamefont {Z.}~\bibnamefont
  {Zhang}}\ and\ \bibinfo {author} {\bibfnamefont {L.-W.}\ \bibnamefont
  {Chen}},\ }\href {\doibase 10.1103/PhysRevC.92.031301} {\bibfield  {journal}
  {\bibinfo  {journal} {Phys. Rev. C}\ }\textbf {\bibinfo {volume} {92}},\
  \bibinfo {pages} {031301(R)} (\bibinfo {year} {2015})}\BibitemShut {NoStop}%
\bibitem [{\citenamefont {Li}\ \emph {et~al.}(2019)\citenamefont {Li},
  \citenamefont {Krastev}, \citenamefont {Wen},\ and\ \citenamefont
  {Zhang}}]{Li:2019}%
  \BibitemOpen
  \bibfield  {author} {\bibinfo {author} {\bibfnamefont {B.-A.}\ \bibnamefont
  {Li}}, \bibinfo {author} {\bibfnamefont {P.~G.}\ \bibnamefont {Krastev}},
  \bibinfo {author} {\bibfnamefont {D.-H.}\ \bibnamefont {Wen}}, \ and\
  \bibinfo {author} {\bibfnamefont {N.-B.}\ \bibnamefont {Zhang}},\ }\href
  {\doibase 10.1140/epja/i2019-12780-8} {\bibfield  {journal} {\bibinfo
  {journal} {Eur. Phys. J. A}\ }\textbf {\bibinfo {volume} {55}},\ \bibinfo
  {pages} {117} (\bibinfo {year} {2019})}\BibitemShut {NoStop}%
\bibitem [{\citenamefont {Russotto}\ \emph {et~al.}(2016)\citenamefont
  {Russotto} \emph {et~al.}}]{Russotto:2016}%
  \BibitemOpen
  \bibfield  {author} {\bibinfo {author} {\bibfnamefont {P.}~\bibnamefont
  {Russotto}} \emph {et~al.},\ }\href {\doibase 10.1103/PhysRevC.94.034608}
  {\bibfield  {journal} {\bibinfo  {journal} {Phys. Rev. C}\ }\textbf {\bibinfo
  {volume} {94}},\ \bibinfo {pages} {034608} (\bibinfo {year}
  {2016})}\BibitemShut {NoStop}%
\bibitem [{prr()}]{prr:supp}%
  \BibitemOpen
  \href@noop {} {}\bibinfo {note} {{See Supplemental Material at
  \url{http://link.aps.org/supplemental/10.1103/PhysRevResearch.2.022033} for a
  table of AFDMC results for the equation of state of pure neutron matter and
  symmetric nuclear matter.}}\BibitemShut {Stop}%
\bibitem [{\citenamefont {Margueron}\ \emph {et~al.}(2018)\citenamefont
  {Margueron}, \citenamefont {Hoffmann~Casali},\ and\ \citenamefont
  {Gulminelli}}]{Margueron:2017}%
  \BibitemOpen
  \bibfield  {author} {\bibinfo {author} {\bibfnamefont {J.}~\bibnamefont
  {Margueron}}, \bibinfo {author} {\bibfnamefont {R.}~\bibnamefont
  {Hoffmann~Casali}}, \ and\ \bibinfo {author} {\bibfnamefont {F.}~\bibnamefont
  {Gulminelli}},\ }\href {\doibase 10.1103/PhysRevC.97.025805} {\bibfield
  {journal} {\bibinfo  {journal} {Phys. Rev. C}\ }\textbf {\bibinfo {volume}
  {97}},\ \bibinfo {pages} {025805} (\bibinfo {year} {2018})}\BibitemShut
  {NoStop}%
\bibitem [{\citenamefont {Gandolfi}\ \emph {et~al.}(2019)\citenamefont
  {Gandolfi}, \citenamefont {Lippuner}, \citenamefont {Steiner}, \citenamefont
  {Tews}, \citenamefont {Du},\ and\ \citenamefont
  {Al-Mamun}}]{Gandolfi:2019zpj}%
  \BibitemOpen
  \bibfield  {author} {\bibinfo {author} {\bibfnamefont {S.}~\bibnamefont
  {Gandolfi}}, \bibinfo {author} {\bibfnamefont {J.}~\bibnamefont {Lippuner}},
  \bibinfo {author} {\bibfnamefont {A.~W.}\ \bibnamefont {Steiner}}, \bibinfo
  {author} {\bibfnamefont {I.}~\bibnamefont {Tews}}, \bibinfo {author}
  {\bibfnamefont {X.}~\bibnamefont {Du}}, \ and\ \bibinfo {author}
  {\bibfnamefont {M.}~\bibnamefont {Al-Mamun}},\ }\href {\doibase
  10.1088/1361-6471/ab29b3} {\bibfield  {journal} {\bibinfo  {journal} {J.
  Phys. G}\ }\textbf {\bibinfo {volume} {46}},\ \bibinfo {pages} {103001}
  (\bibinfo {year} {2019})}\BibitemShut {NoStop}%
\bibitem [{\citenamefont {Epelbaum}\ \emph {et~al.}(2015)\citenamefont
  {Epelbaum}, \citenamefont {Krebs},\ and\ \citenamefont
  {Mei\ss{}ner}}]{Epelbaum:2015}%
  \BibitemOpen
  \bibfield  {author} {\bibinfo {author} {\bibfnamefont {E.}~\bibnamefont
  {Epelbaum}}, \bibinfo {author} {\bibfnamefont {H.}~\bibnamefont {Krebs}}, \
  and\ \bibinfo {author} {\bibfnamefont {U.-G.}\ \bibnamefont {Mei\ss{}ner}},\
  }\href {\doibase 10.1103/PhysRevLett.115.122301} {\bibfield  {journal}
  {\bibinfo  {journal} {Phys. Rev. Lett.}\ }\textbf {\bibinfo {volume} {115}},\
  \bibinfo {pages} {122301} (\bibinfo {year} {2015})}\BibitemShut {NoStop}%
\bibitem [{\citenamefont {Baillot~d'Etivaux}\ \emph {et~al.}(2019)\citenamefont
  {Baillot~d'Etivaux}, \citenamefont {Guillot}, \citenamefont {Margueron},
  \citenamefont {Webb}, \citenamefont {Catelan},\ and\ \citenamefont
  {Reisenegger}}]{Etivaux:2019}%
  \BibitemOpen
  \bibfield  {author} {\bibinfo {author} {\bibfnamefont {N.}~\bibnamefont
  {Baillot~d'Etivaux}}, \bibinfo {author} {\bibfnamefont {S.}~\bibnamefont
  {Guillot}}, \bibinfo {author} {\bibfnamefont {J.}~\bibnamefont {Margueron}},
  \bibinfo {author} {\bibfnamefont {N.}~\bibnamefont {Webb}}, \bibinfo {author}
  {\bibfnamefont {M.}~\bibnamefont {Catelan}}, \ and\ \bibinfo {author}
  {\bibfnamefont {A.}~\bibnamefont {Reisenegger}},\ }\href {\doibase
  10.3847/1538-4357/ab4f6c} {\bibfield  {journal} {\bibinfo  {journal}
  {Astrophys. J.}\ }\textbf {\bibinfo {volume} {887}},\ \bibinfo {pages} {48}
  (\bibinfo {year} {2019})}\BibitemShut {NoStop}%
\bibitem [{\citenamefont {Cai}\ and\ \citenamefont {Chen}(2017)}]{Cai:2014kya}%
  \BibitemOpen
  \bibfield  {author} {\bibinfo {author} {\bibfnamefont {B.-J.}\ \bibnamefont
  {Cai}}\ and\ \bibinfo {author} {\bibfnamefont {L.-W.}\ \bibnamefont {Chen}},\
  }\href {\doibase 10.1007/s41365-017-0329-1} {\bibfield  {journal} {\bibinfo
  {journal} {Nucl. Sci. Tech.}\ }\textbf {\bibinfo {volume} {28}},\ \bibinfo
  {pages} {185} (\bibinfo {year} {2017})}\BibitemShut {NoStop}%
\bibitem [{\citenamefont {Carbone}\ \emph {et~al.}(2014)\citenamefont
  {Carbone}, \citenamefont {Polls}, \citenamefont {Provid{\^e}ncia},
  \citenamefont {Rios},\ and\ \citenamefont {Vida{\~{n}}a}}]{Carbone:2014}%
  \BibitemOpen
  \bibfield  {author} {\bibinfo {author} {\bibfnamefont {A.}~\bibnamefont
  {Carbone}}, \bibinfo {author} {\bibfnamefont {A.}~\bibnamefont {Polls}},
  \bibinfo {author} {\bibfnamefont {C.}~\bibnamefont {Provid{\^e}ncia}},
  \bibinfo {author} {\bibfnamefont {A.}~\bibnamefont {Rios}}, \ and\ \bibinfo
  {author} {\bibfnamefont {I.}~\bibnamefont {Vida{\~{n}}a}},\ }\href {\doibase
  10.1140/epja/i2014-14013-2} {\bibfield  {journal} {\bibinfo  {journal} {Eur.
  Phys. J. A}\ }\textbf {\bibinfo {volume} {50}},\ \bibinfo {pages} {13}
  (\bibinfo {year} {2014})}\BibitemShut {NoStop}%
\bibitem [{\citenamefont {Wellenhofer}\ \emph {et~al.}(2016)\citenamefont
  {Wellenhofer}, \citenamefont {Holt},\ and\ \citenamefont
  {Kaiser}}]{Wellenhofer:2016}%
  \BibitemOpen
  \bibfield  {author} {\bibinfo {author} {\bibfnamefont {C.}~\bibnamefont
  {Wellenhofer}}, \bibinfo {author} {\bibfnamefont {J.~W.}\ \bibnamefont
  {Holt}}, \ and\ \bibinfo {author} {\bibfnamefont {N.}~\bibnamefont
  {Kaiser}},\ }\href {\doibase 10.1103/PhysRevC.93.055802} {\bibfield
  {journal} {\bibinfo  {journal} {Phys. Rev. C}\ }\textbf {\bibinfo {volume}
  {93}},\ \bibinfo {pages} {055802} (\bibinfo {year} {2016})}\BibitemShut
  {NoStop}%
\bibitem [{\citenamefont {Kortelainen}\ \emph {et~al.}(2010)\citenamefont
  {Kortelainen}, \citenamefont {Lesinski}, \citenamefont {More}, \citenamefont
  {Nazarewicz}, \citenamefont {Sarich}, \citenamefont {Schunck}, \citenamefont
  {Stoitsov},\ and\ \citenamefont {Wild}}]{Kortelainen:2010hv}%
  \BibitemOpen
  \bibfield  {author} {\bibinfo {author} {\bibfnamefont {M.}~\bibnamefont
  {Kortelainen}}, \bibinfo {author} {\bibfnamefont {T.}~\bibnamefont
  {Lesinski}}, \bibinfo {author} {\bibfnamefont {J.}~\bibnamefont {More}},
  \bibinfo {author} {\bibfnamefont {W.}~\bibnamefont {Nazarewicz}}, \bibinfo
  {author} {\bibfnamefont {J.}~\bibnamefont {Sarich}}, \bibinfo {author}
  {\bibfnamefont {N.}~\bibnamefont {Schunck}}, \bibinfo {author} {\bibfnamefont
  {M.~V.}\ \bibnamefont {Stoitsov}}, \ and\ \bibinfo {author} {\bibfnamefont
  {S.}~\bibnamefont {Wild}},\ }\href {\doibase 10.1103/PhysRevC.82.024313}
  {\bibfield  {journal} {\bibinfo  {journal} {Phys. Rev. C}\ }\textbf {\bibinfo
  {volume} {82}},\ \bibinfo {pages} {024313} (\bibinfo {year}
  {2010})}\BibitemShut {NoStop}%
\bibitem [{\citenamefont {Chen}\ \emph {et~al.}(2010)\citenamefont {Chen},
  \citenamefont {Ko}, \citenamefont {Li},\ and\ \citenamefont
  {Xu}}]{Chen:2010qx}%
  \BibitemOpen
  \bibfield  {author} {\bibinfo {author} {\bibfnamefont {L.-W.}\ \bibnamefont
  {Chen}}, \bibinfo {author} {\bibfnamefont {C.~M.}\ \bibnamefont {Ko}},
  \bibinfo {author} {\bibfnamefont {B.-A.}\ \bibnamefont {Li}}, \ and\ \bibinfo
  {author} {\bibfnamefont {J.}~\bibnamefont {Xu}},\ }\href {\doibase
  10.1103/PhysRevC.82.024321} {\bibfield  {journal} {\bibinfo  {journal} {Phys.
  Rev. C}\ }\textbf {\bibinfo {volume} {82}},\ \bibinfo {pages} {024321}
  (\bibinfo {year} {2010})}\BibitemShut {NoStop}%
\bibitem [{\citenamefont {Piekarewicz}\ \emph {et~al.}(2012)\citenamefont
  {Piekarewicz}, \citenamefont {Agrawal}, \citenamefont {Colo}, \citenamefont
  {Nazarewicz}, \citenamefont {Paar}, \citenamefont {Reinhard}, \citenamefont
  {Roca-Maza},\ and\ \citenamefont {Vretenar}}]{Piekarewicz:2012pp}%
  \BibitemOpen
  \bibfield  {author} {\bibinfo {author} {\bibfnamefont {J.}~\bibnamefont
  {Piekarewicz}}, \bibinfo {author} {\bibfnamefont {B.~K.}\ \bibnamefont
  {Agrawal}}, \bibinfo {author} {\bibfnamefont {G.}~\bibnamefont {Colo}},
  \bibinfo {author} {\bibfnamefont {W.}~\bibnamefont {Nazarewicz}}, \bibinfo
  {author} {\bibfnamefont {N.}~\bibnamefont {Paar}}, \bibinfo {author}
  {\bibfnamefont {P.~G.}\ \bibnamefont {Reinhard}}, \bibinfo {author}
  {\bibfnamefont {X.}~\bibnamefont {Roca-Maza}}, \ and\ \bibinfo {author}
  {\bibfnamefont {D.}~\bibnamefont {Vretenar}},\ }\href {\doibase
  10.1103/PhysRevC.85.041302} {\bibfield  {journal} {\bibinfo  {journal} {Phys.
  Rev. C}\ }\textbf {\bibinfo {volume} {85}},\ \bibinfo {pages} {041302(R)}
  (\bibinfo {year} {2012})}\BibitemShut {NoStop}%
\bibitem [{\citenamefont {Trippa}\ \emph {et~al.}(2008)\citenamefont {Trippa},
  \citenamefont {Colo},\ and\ \citenamefont {Vigezzi}}]{Trippa:2008gr}%
  \BibitemOpen
  \bibfield  {author} {\bibinfo {author} {\bibfnamefont {L.}~\bibnamefont
  {Trippa}}, \bibinfo {author} {\bibfnamefont {G.}~\bibnamefont {Colo}}, \ and\
  \bibinfo {author} {\bibfnamefont {E.}~\bibnamefont {Vigezzi}},\ }\href
  {\doibase 10.1103/PhysRevC.77.061304} {\bibfield  {journal} {\bibinfo
  {journal} {Phys. Rev. C}\ }\textbf {\bibinfo {volume} {77}},\ \bibinfo
  {pages} {061304(R)} (\bibinfo {year} {2008})}\BibitemShut {NoStop}%
\bibitem [{\citenamefont {Tsang}\ \emph {et~al.}(2009)\citenamefont {Tsang},
  \citenamefont {Zhang}, \citenamefont {Danielewicz}, \citenamefont {Famiano},
  \citenamefont {Li}, \citenamefont {Lynch},\ and\ \citenamefont
  {Steiner}}]{Tsang:2008fd}%
  \BibitemOpen
  \bibfield  {author} {\bibinfo {author} {\bibfnamefont {M.~B.}\ \bibnamefont
  {Tsang}}, \bibinfo {author} {\bibfnamefont {Y.}~\bibnamefont {Zhang}},
  \bibinfo {author} {\bibfnamefont {P.}~\bibnamefont {Danielewicz}}, \bibinfo
  {author} {\bibfnamefont {M.}~\bibnamefont {Famiano}}, \bibinfo {author}
  {\bibfnamefont {Z.}~\bibnamefont {Li}}, \bibinfo {author} {\bibfnamefont
  {W.~G.}\ \bibnamefont {Lynch}}, \ and\ \bibinfo {author} {\bibfnamefont
  {A.~W.}\ \bibnamefont {Steiner}},\ }\href {\doibase
  10.1103/PhysRevLett.102.122701} {\bibfield  {journal} {\bibinfo  {journal}
  {Phys. Rev. Lett.}\ }\textbf {\bibinfo {volume} {102}},\ \bibinfo {pages}
  {122701} (\bibinfo {year} {2009})}\BibitemShut {NoStop}%
\bibitem [{\citenamefont {Danielewicz}\ \emph {et~al.}(2017)\citenamefont
  {Danielewicz}, \citenamefont {Singh},\ and\ \citenamefont
  {Lee}}]{Danielewicz:2016bgb}%
  \BibitemOpen
  \bibfield  {author} {\bibinfo {author} {\bibfnamefont {P.}~\bibnamefont
  {Danielewicz}}, \bibinfo {author} {\bibfnamefont {P.}~\bibnamefont {Singh}},
  \ and\ \bibinfo {author} {\bibfnamefont {J.}~\bibnamefont {Lee}},\ }\href
  {\doibase 10.1016/j.nuclphysa.2016.11.008} {\bibfield  {journal} {\bibinfo
  {journal} {Nucl. Phys. A}\ }\textbf {\bibinfo {volume} {958}},\ \bibinfo
  {pages} {147} (\bibinfo {year} {2017})}\BibitemShut {NoStop}%
\bibitem [{\citenamefont {Tews}\ \emph {et~al.}(2017)\citenamefont {Tews},
  \citenamefont {Lattimer}, \citenamefont {Ohnishi},\ and\ \citenamefont
  {Kolomeitsev}}]{Tews:2017}%
  \BibitemOpen
  \bibfield  {author} {\bibinfo {author} {\bibfnamefont {I.}~\bibnamefont
  {Tews}}, \bibinfo {author} {\bibfnamefont {J.~M.}\ \bibnamefont {Lattimer}},
  \bibinfo {author} {\bibfnamefont {A.}~\bibnamefont {Ohnishi}}, \ and\
  \bibinfo {author} {\bibfnamefont {E.~E.}\ \bibnamefont {Kolomeitsev}},\
  }\href {\doibase 10.3847/1538-4357/aa8db9} {\bibfield  {journal} {\bibinfo
  {journal} {Astrophys. J.}\ }\textbf {\bibinfo {volume} {848}},\ \bibinfo
  {pages} {105} (\bibinfo {year} {2017})}\BibitemShut {NoStop}%
\bibitem [{\citenamefont {Abbott}\ \emph {et~al.}(2018)\citenamefont {Abbott}
  \emph {et~al.}}]{Abbott:2018exr}%
  \BibitemOpen
  \bibfield  {author} {\bibinfo {author} {\bibfnamefont {B.~P.}\ \bibnamefont
  {Abbott}} \emph {et~al.} (\bibinfo {collaboration} {LIGO Scientific
  Collaboration and the Virgo Collaboration}),\ }\href {\doibase
  10.1103/PhysRevLett.121.161101} {\bibfield  {journal} {\bibinfo  {journal}
  {Phys. Rev. Lett.}\ }\textbf {\bibinfo {volume} {121}},\ \bibinfo {pages}
  {161101} (\bibinfo {year} {2018})}\BibitemShut {NoStop}%
\end{thebibliography}
%

\section*{Equation of state tables}
\Cref{tab:eos} reports the AFDMC results for the equation of state of pure neutron matter (PNM) and symmetric nuclear matter (SNM) used in Fig.~1 of the main text.

\begin{table*}[h!]
\centering
\caption[]{Equation of state of PNM and SNM at N$^2$LO for coordinate-space cutoff $R_0=1.0\,\rm fm$ (see Fig.~1 of the main text). Results for the two operator structures $E\mathbbm1$ and $E\tau$ are shown. First error is the statistical Monte Carlo uncertainty. Second error is the uncertainty coming from the truncation of the chiral expansion according to the prescription of Epelbaum \textit{et al.}~\cite{Epelbaum:2015} (see main text for details). Densities are in fm$^{-3}$, energies in MeV$/A$.}
\label{tab:eos}
\begin{ruledtabular}
\begin{tabular}{crrrr}
       & \multicolumn{2}{c}{N$^2$LO $E\mathbbm1$} & \multicolumn{2}{c}{N$^2$LO $E\tau$} \\
$n$ & \multicolumn{1}{c}{PNM} & \multicolumn{1}{c}{SNM} & \multicolumn{1}{c}{PNM} & \multicolumn{1}{c}{SNM} \\ 
\hline
$0.04$ & $ 6.62(0.01)(0.07)$     & \multicolumn{1}{c}{$-$} & $ 6.54(0.03)(0.07)$     & \multicolumn{1}{c}{$-$} \\
$0.08$ & $ 9.91(0.02)(0.31)$     & $ -9.53(0.13)(1.54)$    & $ 9.48(0.05)(0.31)$     & $ -9.65(0.18)(1.58)$    \\
$0.12$ & $12.93(0.09)(0.73)$     & $-11.45(0.15)(3.69)$    & $11.63(0.10)(0.73)$     & $-12.20(0.19)(3.69)$    \\
$0.16$ & $16.12(0.07)(1.34)$     & $-13.07(0.15)(4.55)$    & $12.92(0.21)(1.34)$     & $-13.98(0.14)(4.55)$    \\
$0.20$ & $19.77(0.18)(2.17)$     & $-13.96(0.13)(5.57)$    & $13.15(0.52)(2.17)$     & $-15.43(0.10)(5.82)$    \\
$0.24$ & $23.95(0.23)(3.21)$     & $-13.77(0.10)(6.37)$    & $12.45(1.02)(4.18)$     & $-16.33(0.11)(7.57)$    \\
$0.28$ & $28.70(0.44)(4.47)$     & $-12.97(0.10)(7.88)$    & \multicolumn{1}{c}{$-$} & $-16.84(0.10)(9.78)$    \\
$0.32$ & $34.06(0.39)(5.98)$     & $-10.99(0.10)(8.97)$    & \multicolumn{1}{c}{$-$} & $-17.20(0.07)(12.2)$    \\
$0.40$ & \multicolumn{1}{c}{$-$} & \multicolumn{1}{c}{$-$} & \multicolumn{1}{c}{$-$} & $-17.11(0.13)(-)\phantom{10}$ \\
\end{tabular}
\end{ruledtabular}
\end{table*}

\end{document}